\documentclass[11pt]{article}

\PassOptionsToPackage{nosort}{cite}
\usepackage{soul}
\usepackage[labelfont=normalfont]{subcaption} 
\usepackage{mathrsfs}
\usepackage{dsfont}
\usepackage{amssymb}
\usepackage{float}
\usepackage{enumitem}
\usepackage{slashed}
\usepackage{tikz}
\usepackage[T1]{fontenc}
\usepackage{chet}
\usepackage[font=small,format=hang,labelfont={sf,bf}]{caption}

\usepackage{float}
\newfloat{Newlist}{H}{lol}
\floatname{Newlist}{List}

\usetikzlibrary{decorations.markings, decorations.pathmorphing, arrows, shapes.misc, patterns, intersections}

\newif\ifstartcompletesineup
\newif\ifendcompletesineup
\pgfkeys{
    /pgf/decoration/.cd,
    start up/.is if=startcompletesineup,
    start up=true,
    start up/.default=true,
    start down/.style={/pgf/decoration/start up=false},
    end up/.is if=endcompletesineup,
    end up=true,
    end up/.default=true,
    end down/.style={/pgf/decoration/end up=false}
}
\pgfdeclaredecoration{complete sines}{initial}
{
    \state{initial}[
        width=+0pt,
        next state=upsine,
        persistent precomputation={
            \ifstartcompletesineup
                \pgfkeys{/pgf/decoration automaton/next state=upsine}
                \ifendcompletesineup
                    \pgfmathsetmacro\matchinglength{
                        0.5*\pgfdecoratedinputsegmentlength / (ceil(0.5* \pgfdecoratedinputsegmentlength / \pgfdecorationsegmentlength) )
                    }
                \else
                    \pgfmathsetmacro\matchinglength{
                        0.5 * \pgfdecoratedinputsegmentlength / (ceil(0.5 * \pgfdecoratedinputsegmentlength / \pgfdecorationsegmentlength ) - 0.499)
                    }
                \fi
            \else
                \pgfkeys{/pgf/decoration automaton/next state=downsine}
                \ifendcompletesineup
                    \pgfmathsetmacro\matchinglength{
                        0.5* \pgfdecoratedinputsegmentlength / (ceil(0.5 * \pgfdecoratedinputsegmentlength / \pgfdecorationsegmentlength ) - 0.4999)
                    }
                \else
                    \pgfmathsetmacro\matchinglength{
                        0.5 * \pgfdecoratedinputsegmentlength / (ceil(0.5 * \pgfdecoratedinputsegmentlength / \pgfdecorationsegmentlength ) )
                    }
                \fi
            \fi
            \setlength{\pgfdecorationsegmentlength}{\matchinglength pt}
        }] {}
    \state{downsine}[width=\pgfdecorationsegmentlength,next state=upsine]{
        \pgfpathsine{\pgfpoint{0.5\pgfdecorationsegmentlength}{0.5\pgfdecorationsegmentamplitude}}
        \pgfpathcosine{\pgfpoint{0.5\pgfdecorationsegmentlength}{-0.5\pgfdecorationsegmentamplitude}}
    }
    \state{upsine}[width=\pgfdecorationsegmentlength,next state=downsine]{
        \pgfpathsine{\pgfpoint{0.5\pgfdecorationsegmentlength}{-0.5\pgfdecorationsegmentamplitude}}
        \pgfpathcosine{\pgfpoint{0.5\pgfdecorationsegmentlength}{0.5\pgfdecorationsegmentamplitude}}
}
    \state{final}{}
}

\makeatletter
\newcommand{\namelabel}[1]{%
  \phantomsection
  \renewcommand{\@currentlabel}{#1}
  \label{#1}
}
\makeatother

\definecolor{palecerulean}{rgb}{0.61, 0.77, 0.89}

\tikzset{cross/.style={path picture={
      \draw[black, thick]
            (path picture bounding box.south east) --
            (path picture bounding box.north west)
            (path picture bounding box.south west) --
            (path picture bounding box.north east);}}}

\definecolor{darkblue}{rgb}{0.1,0.1,0.7}
\hypersetup{colorlinks,
           linkcolor={darkblue},
           citecolor={darkblue},
           urlcolor={darkblue},
           breaklinks=true,
           pdftitle={Axions and Superfluidity in Weyl Semimetals},
           pdfdisplaydoctitle
}

\newcommand{\lrpr}{\raise 1ex\hbox{$^\leftrightarrow$} \hspace{-9pt} \partial}
\newcommand{\lrprb}{\raise 1ex\hbox{$^\leftrightarrow$} \hspace{-9pt} \vec{\partial}}
\newcommand{\lpr}{\raise 1ex\hbox{$^\leftarrow$} \hspace{-9pt}\partial}
\newcommand{\rpr}{\raise 1ex\hbox{$^\rightarrow$} \hspace{-9pt}\partial}
\newcommand{\lna}{\raise 1ex\hbox{$^\leftarrow$} \hspace{-9pt}\nabla}
\newcommand{\rna}{\raise 1ex\hbox{$^\rightarrow$} \hspace{-9pt}\nabla}
\newcommand{\lrna}{\raise 1ex\hbox{$^\leftrightarrow$} \hspace{-9.5pt} \nabla}

\newcommand{\lsp}{\hspace{0.5pt}}
\newcommand{\llsp}{\hspace{0.5pt}}

\newcommand{\psib}{\bar{\psi}}

\renewcommand{\ge}{\geqslant}
\renewcommand{\le}{\leqslant}

\renewcommand\a{\alpha}
\renewcommand\b{\beta}
\renewcommand\d{\delta}
\renewcommand\k{\kappa}
\renewcommand\l{\lambda}
\renewcommand\r{\rho}

\renewcommand\t{\tau}
\newcommand\y{\upsilon}
\renewcommand\j{\psi}
\newcommand{\bj}{\bar{\psi}}
\renewcommand\th{\theta}

\renewcommand\c{\chi}

\newcommand\ve{\varepsilon}
\newcommand\e{\epsilon}
\newcommand\g{\gamma}
\newcommand\z{\zeta}
\newcommand\m{\mu}
\newcommand\n{\nu}

\newcommand\p{\pi}
\newcommand\h{\eta}
\newcommand\s{\sigma}
\newcommand\f{\phi}
\newcommand\w{\omega}

\newcommand\vf{\varphi}

\newcommand\tm{\tilde\mu}
\newcommand\tn{\tilde n}

\newcommand\tj{\tilde \jmath}

\renewcommand\L{\Lambda}
\renewcommand\P{\Pi}

\newcommand\G{\Gamma}
\newcommand\F{\Phi}

\renewcommand\S{\Sigma}

\newcommand{\lag}{\langle}
\newcommand{\rag}{\rangle}

\newcommand{\cA}{{\cal A}}
\newcommand{\mA}{{\mathscr A}}
\newcommand{\cD}{{\cal D}}

\newcommand{\cT}{{\cal T}}
\newcommand{\cP} {{\cal P}}

\newcommand{\cS}{\mathcal{S}}

\newcommand{\Tr}{{\rm Tr}}

\newcommand{\bx}{{\bf x}}
\newcommand{\by}{{\bf y}}

\newcommand{\bs}{{\bar \s}}

\newcommand{\jb}{{\bar \j}}

\renewcommand{\vec}{\boldsymbol}
\newcommand{\vx}{\vec x}
\renewcommand{\part}{{\rm part}}

\newcommand{\pq}{p\cdot q}

\newcommand{\sL}{\mathscr{L}}

\newcommand{\be}{\begin{equation}}
\newcommand{\ee}{\end{equation}}
\newcommand{\bes}{\begin{subequations}}
\newcommand{\ees}{\end{subequations}}
\newcommand{\bea}{\begin{eqnarray}}
\newcommand{\eea}{\end{eqnarray}}

\newcommand{\pa}{\partial}

\newcommand{\tr}{\textrm{tr}}

\newcommand{\nn}{\nonumber \\}
\newcommand{\na}{\nabla}

\newcommand{\sdfrac}[2]{\mbox{\small$\displaystyle\frac{#1}{#2}$}}

\def\nbox#1#2{\vcenter{\hrule \hbox{\vrule height#2in
\kern#1in \vrule} \hrule}}
\def\sq{\,\raise.5pt\hbox{$\nbox{.10}{.10}$}\,}

\allowdisplaybreaks
\title{Axions and Superfluidity in Weyl Semimetals}

\author{Emil Mottola,$^{\!a}$ Andrey V.\ Sadofyev,$^{\!b,c}$ and Andreas Stergiou$^{d}$}

\affiliation{$^a$Dept.\ of Physics and Astronomy, Univ.\ of New Mexico, Albuquerque, NM 87131, USA\\
$^b$LIP, Av.\ Prof.\ Gama Pinto, 2, P-1649-003 Lisboa, Portugal\\
$^c$Instituto Galego de F{\'{i}}sica de Altas Enerx{\'{i}}as,
Universidade de Santiago de Compostela,\\\vspace{-3pt} Santiago de Compostela 15782, Spain\\
$^d$Department of Mathematics, King's College London, Strand, London WC2R 2LS, United Kingdom}

\abstract{An effective field theory (EFT) for dynamical axions in Weyl semimetals (WSMs)
is presented. A pseudoscalar axion excitation is predicted in WSMs at sufficiently low temperatures,
independently of the strength of the Weyl fermion self-coupling. For strong fermion self-coupling the
axion is the gapless Goldstone boson of chiral $U(1)^{\text{ch}}$ spontaneous symmetry breaking. For weak
fermion self-coupling an axion is also generated at non-zero chiral density for Weyl nodes displaced in
energy, as a gapless collective mode of correlated fermion pair excitations of the Fermi surface. This
is an explicit example of the extension of Goldstone's theorem to symmetry breaking by the axial anomaly
itself. In both cases the axion is a chiral density wave or phason mode of the superfluid state of the
WSM, and the Weyl fermions form a chiral condensate $\lag \bj \j\rag$ at low temperatures. In
the presence of an applied magnetic field the axion mode becomes gapped, in analogy to the Anderson--Higgs
mechanism in a superconductor. 't Hooft anomaly matching from ultraviolet to infrared scales is
directly verified in the EFT approach. WSMs thus provide an interesting quantum system in which superfluid,
non-Fermi liquid behavior, and a dynamical axion are predicted to follow directly from the axial anomaly
in a consistent EFT that may be tested experimentally. }

\date{October 2023}

\begin{document}

\maketitle

\toc

\section{Introduction: QFT Axions and Emergent Axions in WSMs}

Condensed matter systems can host excitations considered in relativistic quantum field theory (QFT)
and high-energy particle physics, but which have remained unobserved or difficult to study in that setting.
An interesting example spanning these subfields is the emergence of chiral fermion excitations in Weyl
semimetals (WSMs). In a WSM, electronic valence and conduction bands intersect the Fermi level at isolated
points, the Weyl nodes, close to which the quasi-particle spectrum is linear and gapless. The low-lying
excitations are thus Weyl fermions~\cite{XuHasanSci2015, Hosur:2013kxa, rao2016weyl, Armitage2018}, described by two-component
spinors with momentum and spin either aligned or anti-aligned, and hence states of either positive or negative
chirality respectively. The effective Hamiltonian of these gapless Weyl fermion quasi-particles is identical
to that proposed by H.~Weyl almost a century ago~\cite{Weyl:1929}, but so far not observed in high energy
physics, with the Fermi velocity $v_F$ of the WSM  replacing the speed of light $c$ in relativistic QFT.

According to the Nielsen--Ninomiya theorem~\cite{Nielsen:1983rb}, such Weyl nodes must arise in pairs,
with each pair composed of opposite chirality Weyl fermions. Although a WSM often has more than two Weyl
nodes, the minimal and simplest situation is that of a single pair of Weyl nodes, which has also been
encountered~\cite{Wang_2019, Soh_2019}, and upon which we focus for simplicity in this paper.

The emergence of Weyl fermions in WSMs leads to a direct connection with the chiral (also referred to as the
axial, triangle, or Adler--Bell--Jackiw) anomaly~\cite{Adler:1969gk, Bell:1969ts}, the physical consequences of which can
be studied now in a setting very different than the relativistic QFT context in which it was originally found.
In WSMs the chiral anomaly has been shown to be responsible for the quantum anomalous Hall effect and the
chiral magnetic effect~\cite{Son:2012bg,Zyuzin:2012vn,Zyuzin:2012tv,Wang:2012bgb,Goswami:2012db, Vazifeh2013,Basar:2013iaa}. Recent reviews of anomalies and related transport properties in WSMs may be
found in~\cite{Gorbar:2017lnp,Hu2019,Ong:2020ffe}.

Related to the presence of the axial anomaly is the suggested appearance of an axion in
WSMs~\cite{Zyuzin:2012tv, Wang:2012bgb, Goswami:2012db}, {\it i.e.}\ a pseudoscalar `phason' collective
mode that couples linearly to the topological density $\e^{\a\b\m\n} F_{\a\b}F_{\m\n} = 2 F_{\m\n}\widetilde{F}^{\m\n} = 8 \vec E \cdot \vec B$
~\cite{Gooth:2019lmg,NennoGGoothFelserN:2020,Sekine:2020ixs,SaxenaSchm:2021}. In the Standard Model of
particle physics, the strong nuclear interactions are also described by a gauge theory, quantum chromodynamics
(QCD), based on the local non-abelian color group $SU(3)$. Soon after QCD was introduced it was recognized
that the QCD Lagrangian could contain the topological term $\th\,\tr(G_{\m\n}\widetilde{G}^{\m\n})$,
which is the non-abelian generalization of $\th\lsp F_{\m\n}\widetilde{F}^{\m\n}$, with $G_{\m\n}$ the
$SU(3)$ color gauge field strength of QCD, and $\th \in [0, 2\p]$ is an arbitrary phase angle. The problem
is that such a term violates the discrete symmetries of charge conjugation (C) and space reflection parity (P),
whereas the strong nuclear interactions respect these symmetries to a very high degree of accuracy. In fact,
experimental data lead to a bound on $\theta$, $\th \lesssim 10^{-9}$~\cite{Baluni:1979}, consistent with $\th =0$,
whereas QCD itself provides no reason for $\th$ either to be identically zero, or fine tuned to the
required small enough non-zero value.

The axion was introduced as a new particle to solve this strong CP problem of QCD, by promoting the fixed
constant $\th$ angle to a local dynamical field, which could relax to zero or a very small value by its own
dynamics~\cite{PecceiQuinnPRL:1977,PecceiQuinnPRD:1977,Weinberg:1978,Wilczek:1978}. Although the initial
axion proposal was quickly ruled out by experiment, it survives in modified form with much weaker coupling
to known particles, as still the most often considered solution to the strong CP problem of QCD. In addition,
a weakly coupled axion of this kind is a candidate for the dark matter that apparently makes up 25\% of the
total mass-energy in the universe~\cite{ChaEllMar_AxionDM:2022}. This proposed weakly coupled cosmological axion
also has not been detected, after more than four decades of intensive and increasingly sensitive searches,
which continue up to the present~\cite{DuffyvBibber:2009,AxionSearches:2015,AxionDGNV:2020}. Given these null
results to date, from a particle physics perspective, it is worth considering alternatives to a `fundamental'
axion field for a solution to the CP problem. Thus any example where an axion-like mode can be realized as a
collective or emergent phenomenon of quantum many-body physics is interesting, for the possible light it
could shed on the CP `naturalness' problem of the Standard Model. For this, it is important to identify
the requirements and physical mechanism(s) by which an axion can emerge as a collective mode in a realistic
many-body system that can be studied in a laboratory environment, where the parameters of the system
can be varied and controlled. This is just what WSMs provide.

From a  condensed matter perspective, a possible axion excitation in WSMs has been discussed primarily
through the introduction of four-fermion interactions~\cite{Wang:2012bgb} of the Nambu--Jona-Lasinio (NJL)
type~\cite{Nambu:1961tp, Nambu:1961fr}, which induce the spontaneous symmetry breaking (SSB) of
$U(1)^{\text{ch}}$ chiral symmetry. The axion is then the Nambu--Goldstone boson generated by this
SSB of a global symmetry. As a result of this SSB, typically described by the introduction of a scalar
field order parameter developing a non-zero expectation value in its ground state, the Weyl fermions
acquire a mass gap. A preliminary claim of detection of an axionic mode in a WSM has been made
in~\cite{Gooth:2019lmg}, although this awaits confirmation and study of the excitation spectrum
and its detailed properties.

At the same time the axial or chiral anomaly breaks $U(1)^{\text{ch}}$ symmetry explicitly, violating the
Ward--Takahashi (WT) identities of $U(1)^{\text{ch}}$ invariance. Thus it is not clear how this explicit
breaking of a global symmetry by the anomaly is to be reconciled with the usual formulation of Goldstone's
theorem, where the symmetry is assumed to be exact at the microscopic level, and only the ground state of
the theory spontaneously breaks the symmetry. The usual statement of Goldstone's theorem and massless
Goldstone boson depend upon the existence of a bosonic order parameter and degenerate states of the same
energy, related by the symmetry. Neither a scalar order parameter nor the requisite $U(1)^{\text{ch}}$
symmetry are immediately apparent in a fermionic theory possessing an anomaly, which explicitly violates
this very symmetry.

Despite these apparent differences with the usually considered requirements of Goldstone's theorem, it was
shown in a previous paper that a gapless pseudoscalar chiral density wave (CDW) in the anomalous fermion theory
follows from the axial anomaly itself~\cite{MotSad:2021}. The CDW may be understood as a collective excitation
of the Fermi surface and a result of fermion-antifermion (particle-hole) pairing, similar to Cooper pairing
in a superfluid or superconductor. The gapless collective mode is made manifest as a $1/k^2$ massless pole of
the axial anomaly triangle amplitude~\cite{GiaEM:2009}. Thus it appears that the Goldstone phenomenon applies
also the case of anomaly symmetry breaking (ASB), despite---or rather as a direct consequence of---the axial
anomaly itself. Clarifying this state of affairs and applying it to WSMs where the pseudoscalar CDW is an axion
is the first principal aim of this paper. We will show that the familiar description of SSB by a scalar order
parameter and proof of Goldstone's theorem can be extended to the case of anomalous symmetry breaking (ASB).
Indeed both can be derived from the same UV complete QFT in different limits and, remarkably, both lead to
essentially the same low energy effective theory of a relativistic superfluid with an axion excitation.

An interesting related issue is that of decoupling of heavy (gapped) degrees of freedom from the low energy
spectrum, and the status of anomaly matching between the SSB phase, where the fermions become massive, and
the ASB phase, where they were assumed (at least at first) to remain massless. The expectation of 't Hooft
anomaly matching~\cite{tHooft:1979} is that the form and magnitude of the chiral anomaly should be
renormalization group invariant, and hence the same across phases and across widely different distance scales,
even if the fermions develop a mass gap.

Naively a large fermion mass gap might lead one to expect that the fermions and the entire axial anomaly decouple
entirely at low energies or large distances, which would appear to violate 't Hooft anomaly matching
across scales. In the Standard Model this anomaly matching from the short distance scales of perturbative QCD
to the low energy pion dynamics of the confined theory is critical to accounting for the experimentally verified
low energy decay rate $\p^0\to 2\g$, in terms of the quantum numbers of the constituent colored
quarks~\cite{DonGolHol_SM}. Historically this success played an important role in the establishing of QCD as the
theory of the strong interactions. However, a proof of anomaly matching between the short distance, high energy
quark degrees of freedom of QCD and the long distance, low energy theory of mesons in which quarks are
confined and do not appear at all in the low energy spectrum of QCD is necessarily indirect in full
QCD~\cite{tHooft:1979}, because a complete understanding of the mechanism of quark confinement is still lacking.

In the effective field theory (EFT) motivated for application to WSMs of this paper, it turns out that anomaly
matching across energy scales and phases can be checked and verified directly by perturbative methods.
Consequently, this EFT provides a consistent framework from UV to IR in which the fundamental question of
decoupling---or its opposite, 't Hooft anomaly matching---can be addressed systematically. Providing
an explicit example of derivation of low energy superfluid behavior emerging from an EFT, itself
obtained from a UV complete QFT, in which the axial anomaly remains intact is the second principal
goal of this paper.

In order to avoid potential misunderstandings across different sub-fields, let us
emphasize that the EFT approach for WSMs of this paper, which takes full advantage
of the emergent Lorentz symmetry (with Fermi velocity replacing $c$), and Weyl symmetry
(broken only by the axial anomaly) that WSMs enjoy at their Weyl points, can apply only for energies sufficiently close to such a Weyl node. These symmetries are clearly not exact, since the linear dispersion  $E \propto \vec p$ for fermionic excitations near the Weyl node(s) upon which our approach relies, are violated by the non-linear terms in the
fermion spectrum and Fermi arcs at higher energies of the height of the Weyl cones $\L_W$, typically of the order of $100$ meV or more~\cite{XuHasanSci2015,Hosur:2013kxa,rao2016weyl,Armitage2018}. For energies well below this $\L_W$
WSMs provide a laboratory
realization of effective Lorentz invariance where the effects of the axial anomaly of
relativistic QFT and an EFT based on these first principles can be studied systematically,
and the consequences of these approximate symmetries are expected to apply for energies $E \ll \L_W$ at a Weyl node. For simplicity
we focus in this paper on a single isolated Weyl node and neglect the effects of the
non-linear terms suppressed by $E/\L_W$, and other effects due to impurities, etc.

Let us also remark at the outset that another potential source of confusion
is the (at least) three different notions of EFT in common use in the literature:
\begin{enumerate}[label=({\bf\roman*}), align=left, leftmargin=30pt, labelwidth=20pt, itemsep=-2pt]
\item\label{EFTi} The full one-particle irreducible ({\bf 1PI}) quantum effective action $\G[\vf]$
obtained by computing quantum corrections to a classical action $S_{\rm cl}[\f]$, independently of scale;
\item\label{EFTii} The low energy expansion of this non-local 1PI quantum effective action in inverse powers
of the heavy mass $M$ of some quantum field(s) which are completely `integrated out,' to obtain a local
({\bf Wilson}) effective action $\cS_{\text{eff},M}$ applicable to energy scales lower than $M$;
\item\label{EFTiii} The hydrodynamic or {\bf Fluid} effective action $\cS_{\rm Fluid}$ defined only by
symmetries, conservation laws and the equation of state of the system in local thermodynamic equilibrium,
applicable only at macroscopic distance scales, with no reliance upon any specific microscopic theory.
\end{enumerate}

In the case of \ref{EFTi} the generating function of connected correlation functions $W[J]$ is first defined
by the functional integral
\be
\exp\big\{iW[J]\big\}\equiv \int [\cD \f]\lsp \exp\big\{i S_{\text{cl}}[\f] + i\lsp J\circ \f \big\}
\label{Wdef}
\ee
over a set of quantum fields $\{\f_i(x)\}$, governed by the classical action $S_{\text{cl}}[\f]$ in the presence
of the external sources $J_i(x)$. In (\ref{Wdef}) the shorthand notation $J \circ \f$ is employed to denote
$\int d^{\lsp d}x\, J_i(x) \f_i(x)$ in $d$ spacetime dimensions. The 1PI quantum effective action $\G[\vf]$
is then defined as the Legendre transform of $W[J]$~\cite{DeWitt:1964,Weinberg:1995mt},
\be
\G[\vf] \equiv W[J_\vf] - J_\vf \circ \vf\,,
\label{Geffdef}
\ee
evaluated at $J_\vf(x)\equiv J[\vf(x)]$ obtained by solution of the implicit equation
\be
\frac{\d W[J]}{\d J_i(x)}\bigg\vert_{J=J_\vf} = \vf_i(x)\,,
\label{meanfield}
\ee
which must be inverted to find $J_\vf(x)$ and inserted into (\ref{Geffdef}).

The 1PI effective action $\G[\vf]$ so defined is the generating functional of proper vertices, and the $\{\vf_i(x)\}$ are the
background or mean fields of the background field method~\cite{Abbott:1981}. If the original $S_{\text{cl}}$ is renormalizable
and UV complete, its 1PI effective action $\G[\vf]$ and the equations of motion following from it apply at any scale without restriction.
However, in general $\G[\vf]$ formally defined by (\ref{Geffdef}) is a non-local functional of $\vf$, that cannot be calculated
exactly. Restrictions then inevitably arise in the approximations necessary to calculate it. As a familiar example, the one-loop
approximation to $\G[\vf]$ for the $\vf_i$ spacetime constants (not necessarily fundamental fields) are the order parameters of the
effective potential $V_{\rm eff}(\vf)$ at finite temperature or density that can be used to diagnose SSB and the restoration of
symmetry~\cite{KirzLinde:1972}, as well as to prove the Goldstone theorem~\cite{GoldSalamWein:1962,Zumino:1970}. The 1PI effective
potential is therefore closely related to the Landau theory of phase transitions, with parameters
fit to data in condensed  matter applications rather than calculated from first principles of microphysics~\cite{LandauLifStatPhys}.

If some of the quantum fields in (\ref{Wdef}) are massive with large mass gap $M$, one expects that their effects can be neglected
at low energies $E \ll M$, or equivalently distances much larger than $1/M$, as a consequence of {\it decoupling}~\cite{AppCar:1975}.
In that case, such heavy fields may be `integrated out' completely in (\ref{Wdef}), with no sources or mean fields specified
for them.  This procedure is also usually difficult to carry out explicitly, especially when the `fundamental' or UV complete
theory is itself unknown. In that case one relies on the symmetries of the low energy theory and scaling dimensions to expand
the effective action in a power series of local operators of the remaining light fields in ascending powers of $1/M$. A
familiar example of this EFT approach is Chiral Perturbation Theory~\cite{Weinberg:1979,Leut:1994}. Since the separation of
scales specified by $M$ is closely related to the Wilsonian classification of infrared (IR) relevant and irrelevant operators
in statistical and condensed matter physics~\cite{Wilson:1975,Georgi:1993,Alv:2012}, the EFT \ref{EFTii} organized this way in terms
of a heavy mass scale is referred to here as the {\bf Wilson} effective action~\cite{Burgess:2007}.

Finally, at the lowest energy scales when almost no information is available about the underlying microscopic theory or its
fundamental interactions, one can consider a hydrodynamic or {\bf Fluid} EFT \ref{EFTiii} which is based only upon symmetries
and conservation laws, and a specified equilibrium equation of state of the system. Assuming local equilibrium on microscopic
scales ({\it i.e.} on order of the mean free path and smaller) is maintained, small deviations from the equilibrium relations
at the longest wavelength macroscopic scales such as sound waves can be studied~\cite{LandauFluid}.

Generally, each of these three meanings of `effective field theory' is different from the others and the
relationship between them is non-trivial. The remarkable fact following from explicit anomaly matching in
WSMs is that the chiral anomaly admits a consistent description from fundamental principles of QFT connecting
these three approaches, starting from a single renormalizable theory of Weyl fermions coupled to scalars and
electromagnetism in both the SSB and ASB cases, as we shall show, at least at zero temperature and in a
pure sample without defects.

The organization of this paper is as follows. We begin in the next section \ref{Sec:TriAnom} with a review of the axial
or chiral anomaly of fermionic QED$_4$, and the gapless collective boson excitation of
the Fermi surface it implies. In Sec.~\ref{Sec:ChiFlu4} we show how the effective action of the axial anomaly applied to
a WSM with Weyl nodes displaced in energy may be written in a local form in terms of a collective phason field and
recognized as describing a gapless relativistic superfluid chiral sound mode, which is a dynamical propagating axion.
In Sec.~\ref{Sec:Bfield}, we show that this phason mode becomes gapped in the presence of a magnetic field $\vec B$, and
in fact coincides with the massive Schwinger boson of QED$_2$, when its propagation direction is aligned with $\vec B$.
In Sec.~\ref{Sec:NJL} we show how self-interactions of the Weyl fermion modes may be incorporated into the WSM model, by
means of the Nambu--Jona-Lasino model, and Hubbard--Stratonovich transformation, introducing a scalar field $\F$ whose expectation
value for strong enough fermion self-coupling describes the spontaneous breaking of the $U(1)^{\text{ch}}$ chiral symmetry, and
gapping of the Weyl modes. In Sec.~\ref{Sec:fullEFT} we propose a UV renormalizable theory with both fermions and bosons
encompassing both the strong and weak fermion coupling regimes, thus allowing a consistent description of both SSB by the scalar order parameter field and ASB by the axial anomaly collective mode itself at non-zero chiral chemical potential. In
Sec.~\ref{Sec:Perturb}, we study perturbations from equilibrium in this theory and show that the long wavelength
Goldstone sound mode of the scalar $\F$ is the SSB phase satisfes exactly the same dynamical equation of motion as the
axion collective boson of the axial anomaly at finite chiral chemical potential, thus verifying that ASB via the axial
anomaly itself must also lead to a scalar chiral condensate of $\lag \bj \j\rag$ and Goldstone mode.
In Sec.~\ref{Sec:Hooft} the consistency of 't Hooft anomaly matching between the massless Weyl fermion and SSB
cases where the Weyl fermions become gapped is shown explicitly in the EFT across these two cases.
In Sec.~\ref{Sec:Golds} it is shown that Goldstone's theorem can be extended from the more familiar case of
SSB by an explicit scalar order parameter field to the ASB case through the axial anomaly itself at non-zero
$\m_5$, so that the gapless axion collective mode of the previous sections is also a Goldstone boson. Sec.~\ref{Sec:Sum} contains our summary and discussion of results, looking ahead to future work.

There are also three appendices of supplementary material. Appendix \ref{App:Axialanom} contains additional details of
the axial triangle anomaly in QED$_4$, and particularly its infrared apsect, associated spectral sum rule, and
derivation by an unsubtracted Kramers--Kronig dispersion relation. Appendix \ref{App:DimRed} gives the details
of the dimensional reduction of the axial anomaly from $d=4$ to $d=2$ dimensions for a constant, uniform magnetic
field background, and the connection to the Schwinger boson of QED$_2$ described in Sec.~\ref{Sec:Bfield}.
Appendix \ref{App:Intoutfer} gives the details of the integrating out of massive fermions, which provides the
connection between the NJL model of Sec.~\ref{Sec:NJL} and the Wilson effective action \ref{EFTii} of the SSB
phase of a WSM where the Weyl fermion modes acquire a mass gap.

\section{The Axial Anomaly in \texorpdfstring{QED$_\mathbf{4}$}{QED\_4}}
\label{Sec:TriAnom}

The Lagrangian of Dirac fermions $\j$ coupled to electromagnetism with charge strength $e$ is
\be
\sL_f =\bj\lsp\g^\m \big(i \lrpr_\m + e A_\m + b_\m\g^5\big)\j - m\lsp\bj\lsp\j\,,
\label{Lagf}
\ee
where $\lrpr_\m=\frac12(\rpr_\m-\lpr_\m), \jb=\j^\dag\g^0$, and we have allowed for a non-zero fermion mass gap
$m$ and additional coupling to an external axial vector potential $b_\m \equiv A^5_\m$ with unit strength.

The Lagrangian (\ref{Lagf}) is invariant under the local $U(1)^{\rm EM}$ phase transformation
\be
\j \to e^{i \a }\j\,,\qquad \bj \to \bj\lsp e^{-i \a }\,,\qquad eA_\m \to eA_\m + \pa_\m\a
\label{U1EM}
\ee
for any $m$. When $m=0$ (\ref{Lagf}) is also invariant--at the classical level--under the additional
$U(1)^{\rm ch}$ local chiral phase transformation
\be
\j \to e^{i \b\g^5 }\j\,,\qquad \bj \to \bj\lsp e^{i \b \g^5 }\,,\qquad b_\m \to b_\m + \pa_\m\b\,,
\label{U1ch}
\ee
in which the right and left chiralities transform oppositely. By Noether's
theorem these two classical invariances of the action $S_f= \int d^{\lsp 4}x\, \sL_f$ at $m=0$ imply the
conservation of both the electric and axial currents,
\be
J^{\m}= \frac{\d S_f}{\d A_\m} = e\lsp\bj \g^\m \j\,, \qquad\qquad J_5^{\m}= \frac{\d S_f}{\d b_\m}
= \bj \g^\m \g^5\j\,,
\label{NoetherJ}
\ee
with
\be
\hspace{1.15cm} \pa_\m J^\m = 0\,, \hspace{3.35cm} \pa_\m J_5^\m = 2im \bj\g^5\j \qquad \text{(classically)}.
\label{divJ}
\ee
The $m=0$ case thus has an apparent $U(1)^{\text{EM}}\otimes U(1)^{\text{ch}}$ symmetry.

\begin{figure}[ht]
\centering
\begin{tikzpicture}[decoration={snake, segment length=2.06mm, amplitude=0.7mm}]
    \tikzstyle{arr}=[decoration={markings,mark=at position 1 with
      {\arrow[scale=1.5]{>}}}, postaction={decorate}];
    \tikzstyle{arrmid}=[decoration={markings,mark=at position 0.5 with
      {\arrow[scale=1.5]{>}}}, postaction={decorate}]
    \draw[arrmid] (0,0)--(2,1.5);
    \draw[arrmid] (2,1.5)--(2,-1.5);
    \draw[arrmid] (2,-1.5)--(0,0);
    \draw[decorate] (3.5,1.5)--(2,1.5);
    \draw[decorate] (3.5,-1.5)--(2,-1.5);
    \draw[arr] (-1.5,0)--(-1,0) node[above, midway] {$k$};
    \draw[arr] (2.5,1.8)--(3,1.8) node[above, midway] {$p$};
    \draw[arr] (2.5,-1.8)--(3,-1.8) node[below, midway] {$q$};
    \filldraw[cross,fill=white,thick] (0,0) circle (4pt) node[left=2pt] {$J^\m_5$};
    \filldraw[fill=black] (2,1.5) circle (1.5pt) node[above=2pt] {$J^\alpha$};
    \filldraw[fill=black] (2,-1.5) circle (1.5pt) node[below=2pt] {$J^\beta$};
\end{tikzpicture}
\caption{The one-loop axial anomaly $\lag J_5^\m J^\a J^\b\rag$ triangle diagram, which is $\G_5^{\m\a\b}(p,q)$
of (\ref{Amp5mom}) in momentum space. The solid lines represent the propagators of Dirac fermions
and the wavy lines external photon legs, which carry off four-momenta $p$ and $q$ from the electromagnetic
current vertices $J^\a$ and $J^\b$. The incoming four-momentum at the axial current vertex $J^\m_5$
is $k=p+q$ by momentum conservation.}
\label{Fig:J5JJ}
\end{figure}
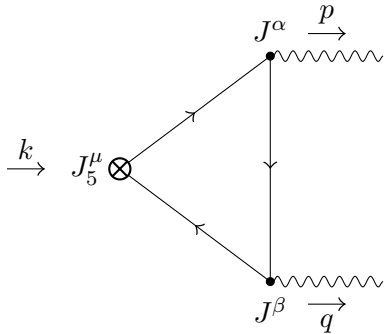

As is well-known, this apparent larger classical symmetry at $m=0$ does not survive in the quantum theory,
since it turns out to be impossible to simultaneously satisfy the requirements of Lorentz invariance,
$U(1)^{\text{EM}}$ gauge invariance and $U(1)^{\text{ch}}$ chiral invariance at the quantum many-particle
level~\cite{Adler:1969gk,treiman2015lectures,Bertlmann:2000}. This conflict of symmetries first appears
at the one-loop level of the triangle diagram of Fig.~\ref{Fig:J5JJ}.

The triangle diagram $\G_5^{\m\a\b}(p,q)$ is naively
ultraviolet (UV) linearly divergent at large loop momenta or short distances, and hence is {\it a priori undefined} and in need of some further prescription. As reviewed in Appendix \ref{App:Axialanom}, it is rendered finite and well-defined by the requirements of:
\begin{enumerate}[nosep, leftmargin=1cm, label=(\alph*)]
\item\label{inva} Lorentz and parity/time reversal invariance of the vacuum or ground state (\ref{Amp5mom});\label{reqi}
\item\label{invb} Electric charge current conservation (\ref{taucons}); \label{reqii}
\item\label{invc} Bose symmetry (\ref{chibose}) under interchange of the photon legs;\label{reqiii}
\item\label{invd} Real part obtained from imaginary part by unsubtracted dispersion (Kramers--Kronig) relation.
\end{enumerate}
With these four requirements $\G_5^{\m\a\b}(p,q)$ does {\it not}
satisfy axial or chiral invariance (\ref{divJ}) in the massless fermion limit $m\to 0$.

In fact, contraction of $\G_5^{\m\a\b}(p,q)$ defined by requirements \ref{inva}-\ref{invd} with the external
momentum $k_{\m} = p_\m+ q_\m$ entering at the axial vector vertex gives the well-defined result
\be
k_\m \, \G_5^{\m\a\b}(p,q) = \cA \, \y^{\a\b}(p,q)\,,
\label{axanom}
\ee
where
\be
\y^{\a\b}(p,q) \equiv \e^{\a\b \r\s} p_{\r}q_{\s} =  \y^{\b\a}(q,p)
\label{upsdef}
\ee
and $\cA$ is most conveniently presented as an integral over Feynman parameters $x,y$ in the form
\be
\cA (k^2; p^2,q^2;m^2) =\frac{e^2}{2\pi^2}\left(1 - 2\lsp m^2\,\int_0^1 dx\int_0^{1-x} dy\  \frac{1}{D}\right),
\label{axdiv}
\ee
where the denominator $D$ in (\ref{axdiv}) is given by (\ref{denom}). The second term proportional to $m^2$
is what would be expected from the axial vector divergence $\pa_\m J_5^\m = 2im \bj\g^5\j$ (\ref{divJ}) following
from use of the on-shell Dirac equation for fermions of mass $m$, {\it cf.} (\ref{Am})~\cite{Horejsi:1985}.
The first term in (\ref{axdiv}), namely
\be
\cA(k^2; p^2,q^2;m^2\!=\!0) = \frac{e^2}{2 \p^2}\,,
\label{A0}
\ee
which survives even in the case of massless fermions $m\to 0$, is the finite axial or chiral anomaly, since it
violates the expectation of the classical divergence of $\pa_\m J_5^\m$ (\ref{divJ}) vanishing at $m=0$.

The result (\ref{axdiv})-(\ref{A0}) is equivalent to
\be
\pa_\m \lag J^\m_5\rag\Big\vert_{m=0} \equiv \mA_4 =\frac{e^2}{16\p^2}\, \e^{\a\b\m\n}F_{\a\b} F_{\m\n}
= \frac{\a}{2\p} F_{\m\n}\widetilde{F}^{\m\n} = \frac{2\a}{\p}\, \vec E \cdot \vec B
\label{divJ5}
\ee
in position space, in the presence of external electric $\vec E$ and magnetic $\vec B$ fields. Here
$\widetilde{F}^{\m\n} \equiv \frac{1}{2}\lsp \e^{\a\b\m\n} F_{\a\b}$ is the dual of $F_{\m\n}$, and
$\a = e^2/4\p \simeq 1/137.036$ is the fine structure constant.

The determination of $\G_5^{\m\a\b}$ and (\ref{axdiv}) solely by the conditions
\ref{inva}-\ref{invd} above shows that these symmetry requirements are all that are necessary and sufficient to determine the anomaly (\ref{divJ5}),
which is finite and independent of any UV divergences or renormalization.  Thus the axial anomaly may be viewed
as a low energy or {\it macroscopic} feature of QFT, which is why it is relevant for EFT treatments of both
particle physics and condensed matter systems such as WSMs.

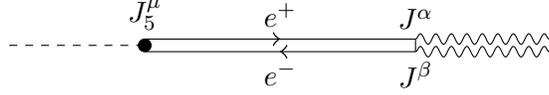
\begin{figure}[ht]
\centering
\begin{tikzpicture}[scale=1.2,decoration={snake, segment length=2.06mm, amplitude=0.7mm}]
    \tikzstyle{arr}=[decoration={markings,mark=at position 1 with {\arrow[scale=1.5]{>}}}, postaction={decorate}];
    \tikzstyle{arrmid}=[decoration={markings,mark=at position 0.5 with
      {\arrow[scale=1.5]{>}}}, postaction={decorate}]
    \draw[dashed] (0,0)--(1.5,0);
    \draw[arrmid] (1.5,0.07)--(4.5,0.07) node[above=1pt, midway] {$e^+$};
    \draw[arrmid] (4.5,-0.07)--(1.5,-0.07) node[below, midway] {$e^-$};
    \filldraw[fill=black] (1.5,0) circle (2pt) node[above=2pt] {$J^\m_5$};
    \draw (4.5,-0.07)--(4.5,0.07);
    \draw[decorate] (4.5,0.07)--(6,0.07) node[above, at start] {$J^\alpha$};
    \draw[decorate] (4.5,-0.07)--(6,-0.07) node[below, at start] {$J^\beta$};
\end{tikzpicture}
\caption{The $e^+e^-$ co-linear massless fermion pair state of opposite helicities from
the triangle diagram of Fig.~\ref{Fig:J5JJ} represented as a single massless effective boson.}
\label{Fig:epair}
\end{figure}

The triangle amplitude $\G_5^{\m\a\b}(p,q)$ for massless fermions may be decomposed as
\be
\G_5^{\m\a\b}(p,q) = \frac{2\a}{\p} \,\frac{k^\m\!}{k^2\!}\ \e^{\a\b\r\s} p_{\r}q_{\s}+\G_{5\,\perp}^{\m\a\b}(p,q)\,,
\label{anompole}
\ee
where $\G_{5\,\perp}^{\m\a\b}$ is transverse, $k_\m \G_{5\,\perp}^{\m\a\b}(p,q)= 0$, and hence non-anomalous,
while the first term in (\ref{anompole}) which is responsible for the anomaly contains a massless $1/k^2$
pole. Whereas the transverse part may (and does) receive all manner of radiative corrections from higher order
processes, the longitudinal anomalous pole contribution in (\ref{anompole}) is protected from such
corrections by the Adler--Bardeen theorem~\cite{AdlerBard:1969}. This remarkable fact can be understood
to be a result of the topological character of the anomaly equation (\ref{divJ5})~\cite{Zumino:1984,WittJackTreiZumi:1985},
and another indication of its long distance properties. The topologically protected massless pole at
$k^2=0$ in the anomaly term describes a two-particle fermion/anti-fermion intermediate state in the triangle
diagram, with this fermionic pair propagating coherently and co-linearly at the speed of
light~\cite{ColemanGrossman}, {\it cf.} Fig.~\ref{Fig:epair}. This fermion pair state of opposite
helicities is just that of a massless {\it boson}~\cite{GiaEM:2009}.

This gapless bosonic mode is a collective excitation of the Fermi--Dirac sea, revealed first in one spatial
dimension in the Schwinger model of massless QED$_2$~\cite{Schwinger:1962tp}, and in fermionic condensed
matter systems as Luttinger liquids~\cite{Luttinger:1963, MattisLieb:1965, Haldane:1981JP}. Similar Luttinger
liquid-like behavior has been conjectured in higher dimensional fermion systems at low
temperatures~\cite{Haldane:2005, HouMar:1993, CasFrad:1994}. It is the $1/k^2$ anomaly pole in
(\ref{anompole}) that provides the theoretical basis for this conjecture to be physically realized in WSMs,
with the Fermi velocity $v_F$ replacing the speed of light $c$, and the fermion pairing to be
analogous to the Cooper pairs of a superfluid condensate~\cite{MotSad:2021}.

The bosonic excitation arising directly from the axial anomaly itself provides a new realization of Goldstone's
theorem, anomaly symmetry breaking (ASB), discussed in Sec.~\ref{Sec:Golds}, without (so far) any explicit
reference to a bosonic order parameter expectation value violating the $U(1)^{\text{ch}}$ symmetry, as in the
usual case of SSB. We shall see that despite the fact that the axial anomaly modifies the WT identities of the
theory whereas SSB preserves them, the two apparently distinct situations of ASB and SSB are very closely related
and can be described by the same low energy EFT, as a superfluid with a Goldstone sound mode that is an
axion excitation in a WSM.

\section{Anomaly Effective Action and Superfluid EFT of Weyl Nodes Displaced in Energy}
\label{Sec:ChiFlu4}

Since the axial current $J^\m_5$ is the variation of the action with respect to the external
axial potential $b_\m$, the longitudinal projection of (\ref{anompole}) is the variation of the non-local 1PI effective action \ref{EFTi}  \cite{GiaEM:2009,Smailagic:2000,Blaschke:2014}
\be
\G_{\rm anom}^{\rm NL}[A, b] = \int d^{\lsp 4}x \int d^{\lsp 4} y\,b_\m(x)\, \lsp\pa_x^\m\, \sq^{-1}_{xy}\,\mA_4(y)
=\frac{\a}{2\pi} \int d^{\lsp 4}x \int d^{\lsp 4}y\, \big[b_\m\lsp\pa^\m\big]_x \sq^{-1}_{xy}\,
\big[F_{\a\b}\lsp \widetilde{F}^{\a\b}\big]_y
\label{SNL4}
\ee
in position space, where $\sq^{-1}_{xy}= \frac{1\,}{4 \pi^2}\, (x-y)^{-2}$ denotes the massless scalar propagator
inverse of the wave operator $\sq= \pa^\m\pa_\m =-\pa_t^2 + \na^2$ in $3+1$ spacetime dimensions and $\mA_4$ the
axial anomaly given by (\ref{divJ5}). The effective action (\ref{SNL4}) is non-local, although its variation with
respect to $b_\m$ which gives $\lag J^\m_5\rag$, and subsequent acting with $\pa/\pa x^\m$, reproduces the local result
(\ref{divJ5}).

The non-local 1PI effective action (\ref{SNL4}) can also be expressed in a local form
\be
\cS_{\rm anom}[\h; A,b]= \int d^{\lsp 4}x\, \Big[\big(\pa_\m\h + b_\m\big)\lsp J_5^\m + \h\lsp\mA_4\Big]
\label{Sanom}
\ee
upon the introduction of the local pseudoscalar potential $\h$. The variation of (\ref{Sanom}) with respect
to $\h$ reproduces the anomaly (\ref{divJ5}), while the variation of (\ref{Sanom}) with respect to the hydrodynamic
current $J_5^\m$ produces the constraint $\pa_\m\h + b_\m= 0$. Solving this constraint for $\h$ gives $\h=-\sq^{-1} \pa^\m b_\m$,
and substituting this back into (\ref{Sanom}) reproduces the non-local action (\ref{SNL4}) with its massless pole,
after integration by parts. This variational method with respect to a current in which no reference to the underlying
QFT variables $\j$ is made, is employed for effective actions of fluid hydrodynamics \ref{EFTiii}, in
which context $\h$ is called a \emph{Clebsch potential}~\cite{JackNair:2004}.

Note that apart from the anomaly term $\h\lsp\mA_4$, (\ref{Sanom}) depends upon the linear combination
$\pa_\m \h + b_\m$, which is invariant under $\h \to \h - \b, b_\m \to b_\m + \pa_\m \b$. This is
a suggestive remnant of the local $U(1)^{\text{ch}}$ symmetry (\ref{U1ch}), although the fermions themselves
do not appear explicitly in the anomaly effective action (\ref{Sanom}). The fluid form has only bosonic
variables, and a fluid variational principle in which only the full current $J^\m_5$ is varied, not
its fermionic constituents. Thus it is not immediately apparent from (\ref{Sanom}) to what complex
bosonic variable $\F$ the phase $\h$ corresponds, or how its magnitude $|\F|$ is related to the underlying
fermion QFT.

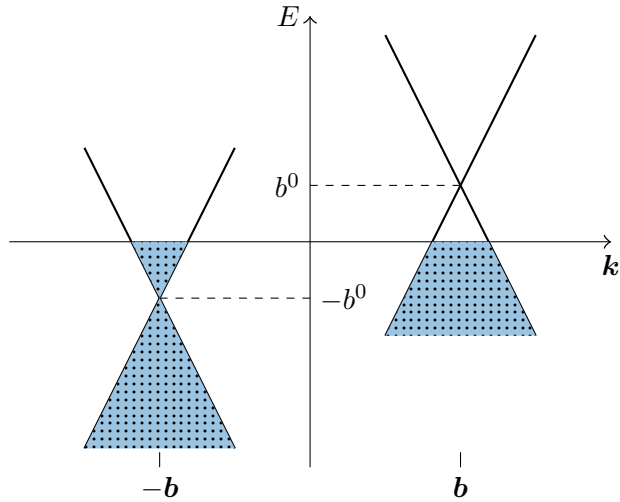
\begin{figure}[ht]
\centering
\begin{tikzpicture}
  \tikzstyle{arr}=[decoration={markings,mark=at position 1 with {\arrow[scale=1.5]{>}}}, postaction={decorate}];
  \draw[name path=xaxis, arr] (-4,0)--(4,0) node[below=1pt] {$\vec{k}$};
  \draw[arr] (0,-3)--(0,3) node[left] {$E$};
  \draw[thick, name path=LNWSE] (-1,-2.75)--(-3,1.25);
  \draw[thick, name path=LNESW] (-3,-2.75)--(-1,1.25);
  \draw[thin, dashed] (-2,-0.75)--(0,-0.75) node[right] {$-b^0$};
  \fill[fill=palecerulean, fill opacity=0.8] (-1,-2.75)--(-2,-0.75)--(-3,-2.75)--cycle;
  \fill[pattern=dots, pattern color=black] (-1,-2.75)--(-2,-0.75)--(-3,-2.75)--cycle;
  \path[name intersections={of=LNWSE and xaxis, by=L1}];
  \path[name intersections={of=LNESW and xaxis, by=L2}];
  \fill[fill=palecerulean, fill opacity=0.8] (-2,-0.75)--(L1)--(L2)--cycle;
  \fill[pattern=dots, pattern color=black] (-2,-0.75)--(L1)--(L2)--cycle;
  \draw[thin] (-2,-2.8)--(-2,-3) node[below] {$-\vec{b}$};
  \draw[thick, name path=RNESW] (1,-1.25)--(3,2.75);
  \draw[thick, name path=RNWSE] (3,-1.25)--(1,2.75);
  \draw[thin, dashed] (2,0.75)--(0,0.75) node[left] {$b^0$};
  \path[name intersections={of=RNESW and xaxis, by=R1}];
  \path[name intersections={of=RNWSE and xaxis, by=R2}];
  \fill[fill=palecerulean, fill opacity=0.8] (1,-1.25)--(R1)--(R2)--(3,-1.25)--cycle;
  \fill[pattern=dots, pattern color=black] (1,-1.25)--(R1)--(R2)--(3,-1.25)--cycle;
  \draw[thin] (2,-2.8)--(2,-3) node[below] {$\vec{b}$};
\end{tikzpicture}
\caption{Two Weyl cones separated by $(2b^0,2\vec{b})$ in momentum space. Here we assume zero charge
chemical potential, so the Weyl cones are filled up to the Fermi energy, as indicated by the fill pattern.
We also neglect the effects of non-linear interactions at energies $\L_W$ far from the Weyl nodes, so that
the EFT description applies for energy excitations and $|2b^0| \ll \L_W$.}
\label{Fig:WeylCones}
\end{figure}

On the constraint $\pa_\m\h + b_\m= 0$, the $\h\lsp \mA_4$ term alone in (\ref{Sanom}) is the form of the effective
action of the axial anomaly that can be shown to be responsible for the anomalous Hall conductance, as well as the
chiral magnetic and chiral separation effects~\cite{Zyuzin:2012tv, MotSad:2021}. Further, if
$2\lsp b^\m =(2\lsp b^0, 2\lsp b^i)$ is the difference in the energies and momenta of the two Weyl nodes
of a WSM respectively, illustrated in Fig.~\ref{Fig:WeylCones}, then
$\h = -x^\m b_\m =  t\lsp b^0 -\lsp {\vec x}\cdot \vec{b}$ is the chiral phase obtained by a fixed chiral
rotation on the fermions through Fujikawa's method~\cite{Fujikawa:1979ay,Fujikawa:1980eg}, and (\ref{Sanom}),
giving the effective action of the axial anomaly for such WSMs in~\cite{Hosur:2013kxa, Zyuzin:2012tv, Goswami:2012db}.

In the WSM literature, the phase $\th = -2\h$ multiplying the axial anomaly $\mA_4$ in (\ref{Sanom}) is called
an `axion.' However, in QFT the axion of \cite{Weinberg:1978,Wilczek:1978} is a {\it propagating} pseudoscalar
field, not simply a chiral rotation angle, of a fixed $b_\m$ axial potential characterizing Weyl nodes in
their equilibrium configuration. To obtain a true propagating axion in a WSM the two Weyl nodes of
Fig.~\ref{Fig:WeylCones} must be allowed to fluctuate around their equilibrium values, and a kinetic
term for $\h$ must be added. Such a kinetic term may just be assumed to be present in the EFT,
as IR relevant terms in the Wilsonian sense \ref{EFTii}, {\it e.g.} \cite{li2010dynamical, Wang:2012bgb, Curtis:2022qis}.
However the origin of this kinetic term is not specified in that case and its coefficient must be
treated as an unknown free parameter that can only be fixed by experiment in such an approach.

On the other hand, continuing with the {\bf Fluid} description \ref{EFTiii} of free massless fermions with
displaced Fermi levels for left- and right-handed Weyl fermions automatically provides just such a
kinetic term for small deviations from equilibrium, with coefficient fixed by the equilibrium equation of state,
once the energy density $\ve(n_5)$ of the fermions as a general function of the chiral number density
$n_5$ is taken into account. Adding this non-anomalous $-\ve(n_5)$ term to the local anomaly effective
action (\ref{Sanom}) results in the fluid effective action
\be
\cS_{\rm Fluid} = \cS_{\rm anom}[\h; A,b] - \int d^{\lsp 4} x\, \ve(n_5) =
\int d^{\lsp 4}x \Big[\big(\pa_\m\h + b_\m\big)J_5^\m + \h\lsp \mA_4 - \ve (n_5)\Big]
\label{Sfer}
\ee
for the massless fermion system at finite $n_5$. The variation of this effective action with respect to
$J_5^\m$ is now non-trivial and leads to
\be
\pa_\m\h  +  b_\m - \left(\!\frac{d\ve}{d n_5}\!\right)\left(\! \frac{dn_5}{d J_5^\m}\!\right)=0 \quad \Rightarrow
\quad  J_5^\m= -\frac{n_5}{\m_5}\,\big(\pa^\m\h +  b^\m\big)
\label{J5eta4}
\ee
instead of the constraint $\pa_\m\h  +  b_\m = 0$, and where the Lorentz frame-invariant definition of
$n_5^2 = -J_5^\m J_{5\,\m}$ has been used. The quantity $\m_5$, likewise invariantly defined by
\be
\m_5 = \frac{d\ve}{d n_5}\,, \qquad \m_5^2 = - \big(\pa^\m\h +  b^\m\big)\big(\pa_\m\h +  b_{\m}\big)\,,
\label{mu5}
\ee
carries the interpretation of the axial chemical potential of the chiral effective fluid. Using these
definitions, (\ref{Sfer}) can be expressed in the form
\be
\cS_{\rm eff,fluid} = \int d^{\lsp 4}x\, \big(\m_5 n_5 - \ve + \h\lsp \mA_4\big)
= \int d^{\lsp 4}x\, \big(P + \h\lsp \mA_4\big)\,,
\label{Sfluid}
\ee
where $P=\m_5 n_5 - \ve$ is the equilibrium pressure of the fluid at zero temperature.

For a WSM whose two Weyl nodes of opposite chirality are displaced in energy and momentum space by $2\lsp b^0$ and $2\lsp b_i$
respectively, as in Fig.~\ref{Fig:WeylCones}, in the rest frame of the material in equilibrium all time derivatives vanish,
whereas $J^0_5 = n_5$ is non-vanishing. Thus when the non-zero $\ve(n_5)$ and $\m_5$ of the fluid in (\ref{Sfer}) are taken
into account, we have
\begin{subequations}
\begin{align}
\bar \h = - \lsp {\vec x}\cdot \vec{b}&\,,\qquad\qquad \dot{\bar\h} =0\,, \label{etabar}\\
\m_5 = b_0&\,,\qquad\qquad \bar{J}^0_5 = n_5\,,
\end{align}
\end{subequations}
and the electric current
\be
J^\m=\frac{2\a}{\p}\lsp\widetilde{F}^{\m\n}\lsp\pa_\n\h\quad\Rightarrow\quad
\bar{\!\vec{J}}=\frac{2\a}{\p}\lsp \vec{B} \lsp\dot{\bar \h} =0
\label{Janom}
\ee
and axial current $\bar{\!\vec{J}}_{\!5}=0$ vanish in equilibrium from (\ref{J5eta4}) and (\ref{etabar}),
including also in the presence of a background magnetic field $\vec B$. In equilibrium with $\vec E = 0$, the
anomaly $\mA_4$ also vanishes, and $\pa_\m \bar J_5^\m= 0$. Note that as a result of \eqref{J5eta4} and using $\bar{J}^0_5=n_5$ and $\bar{\!\vec{J}}_{\!5}=0$ in equilibrium, we find $\dot{\bar{\eta}}=\mu_5-b_0$, and the anomalous electric current is zero in equilibrium, consistent with the literature \cite{Landsteiner:2016led}.

Now since $\h$ is a dynamical field in the effective fluid description and $J^\m_5$ depends on
$\h$ through (\ref{J5eta4}), small fluctuations $\d \h (t, \bx)$ from equilibrium can be considered
and these will generate small fluctuations in $\m_5$ and $J^\m_5$, according to
\begin{subequations}
\begin{align}
\d \m_5 & = \d \dot \h\,, \\
\d J_5^0 & = \d n_5 = \frac{\d n_5}{\d \m_5}\, \d \m_5 = \frac{\d n_5}{\d \m_5}\, \d \dot \h\,,  \label{deln5}\\
\d J^i_5 & = - \frac{n_5}{\m_5}\, \na^i \, \d \h \,,
\end{align}
\label{J5eta}\end{subequations}
for the chiral current components and their departure from equilibrium. To first order in these departures
from equilibrium the variation in (\ref{deln5}) may be replaced by its value from the equilibrium equation of state,
\be
\frac{\d n_5}{\d \m_5} \simeq \frac{d n_5}{d \m_5}\,,
\ee
where we drop overbars on the equilibrium values of $\m_5, n_5$, which are independent of $t,\vec x$.
Now using the definition of the sound speed in the fluid
\be
v_s^2 = \frac{dp}{d\ve}=\frac{n_5\, d\m_5}{\m_5\, dn_5} =
\frac{n_5}{\m_5}\left(\frac{d n_5}{d\m_5}\right)^{\!-1}\,,
\label{vels4}
\ee
the axial anomaly equation $\pa_\m (\d J^\m_5) = \d \mA_4$ for the linear variations in (\ref{J5eta})
may be expressed in the form
\be
\pa_\m(\d J^\m_5)= \frac{d n_5}{d \m_5} \left(\frac{\pa^2}{\pa t^2} - v_s^2\, \vec\na^2\right) \d\h
 = \frac{\a}{2\pi} \,\d \big( F_{\m\n} \widetilde{F}^{\m\n}\big) \,.
\label{etaeom4}
\ee
This follows also from (\ref{Sfer}), expanded to quadratic order in the departures from equilibrium, {\it i.e.}
\begin{align}
\cS^{(2)}_{\rm eff, fluid} & \simeq
\int d^{\lsp 4} x\, \bigg[ \frac{1}{2} \frac{d n_5}{d \m_5}\, \d\h \left(\!-\frac{\pa^2}{\pa t^2} +
v_s^2\, \vec\na^2\right) \d\h +\d\h\, \d\mA_4\bigg]\,.
\label{S2fluid}
\end{align}
If the anomaly term $\d\mA_4$ is assumed to be of the same order of small variation $\d\h$ from equilibrium,
the coupling of $\d\h$ to electromagnetism would need to be considered. However since $\mA_4 \sim \vec E \cdot \vec B$
is quadratic in the EM fields, this will only be the case if there is a non-zero background field of either
$\vec E$ or $\vec B$. The latter is considered in the next section. Otherwise if both $\vec E$ and $\vec B$
are flucutations from the vacuum of $\vec E = \vec B =0$, the last term in (\ref{S2fluid}) is third order in
fluctuations and the EM influence on $\d\h$ can be neglected. This last term in (\ref{S2fluid})
will nevertheless generate an electric current as in (\ref{Janom}) that acts as source for Maxwell's equations.

Eq.~(\ref{etaeom4}) shows that the effective action of the axial anomaly, expressed in the hydrodynamic
fluid form (\ref{Sanom})-(\ref{Sfer}), predicts the existence of a gapless pseudoscalar chiral density wave (CDW)
that couples linearly to $F\widetilde F$ and therefore may be identified as an axion for a WSM arising from displaced
Weyl nodes, as in Fig.~\ref{Fig:WeylCones}, that fluctuate from their equilibrium value. This axionic CDW
arises from the axial anomaly itself, independently of direct fermion self-interactions. For the specific
example of free Weyl fermions we have the zero temperature equilibrium relations
\bes
\bea
&&n_5 = \left(J_5^\m J_{5\,\m}\right)^{\!\frac{1}{2}}=
2\! \int_{|\vec p| \le \m_5}  \frac{d^3 \vec p}{(2 \pi)^3} = \frac{\m_5^3}{3 \pi^2}\,, \\
&&\ve (n_5) =  2\! \int_{|\vec p| \le \m_5}  \frac{d^3 \vec p}{(2 \pi)^3}\,  |\vec p| = \frac{\m_5^4}{4 \pi^2}
= \sdfrac{3}{4} \,(3 \pi^2)^{\frac{1}{3}}\, n_5^{\frac{4}{3}}\,,\\
&&P(\m_5) = \m_5 n_5 - \ve =\frac{\m_5^4}{12 \p^2} =\sdfrac{\ve}{3}\,,\qquad\quad \frac{dP}{d\m_5} = n_5
\eea
\label{ener4}\ees
and the speed of the CDW, $v_s$ in this specific case of free fermions with Weyl nodes displaced
in energy and momentum, is given by
\be
v_s^2\big\vert_{\text{free}} = \sdfrac{1}{3} \lsp v_F^2\,,
\label{vCDW}
\ee
where we have reinserted the Fermi velocity $v_F$ which plays the role of the speed of light in the
relativistic (gapless) Weyl fermion Hamiltonian.

These simple considerations following from the axial anomaly itself in its effective fluid description
(\ref{Sfluid}) predict the existence of a true dynamical and propagating collective axion excitation of
a WSM with a pair of Weyl nodes displaced in energy. In Sec.~\ref{Sec:Golds} we shall show that the appearance
of this gapless mode of a WSM can be understood as a consequence of an extension of Goldstone's theorem to
the case of anomaly symmetry breaking. Physically, it is a collective mode arising from small
perturbations of the Fermi surface of a WSM, with the fermion pairing arising through electromagnetism and
the axial anomaly itself, as in Fig.~\ref{Fig:epair}. At zero temperature and in the absence of defects this
CDW axionic mode is dissipationless, and the effective action (\ref{Sfer}) of the WSM fluid is that of
a relativistic superfluid~\cite{Schmitt:2014eka}.

\section{The Axion for WSMs in a Constant Uniform Magnetic Field: Dimensional Reduction}
\label{Sec:Bfield}

As a second example of the application of the hydrodynamic EFT to a WSM, consider a WSM,
for free or very weakly self-interacting fermions, and with no offset in energy of the Weyl nodes,
{\it i.e.}\ $b^0 =0$, but placed in a constant, uniform magnetic field $\vec B= B \bf \hat x$.
In this case the triangle diagram and vertex  $\G_5^{\m\a\b}$ reduces to its longitudinal component
only and the effective action restricted to excitations along $\vec B$ becomes exact~\cite{MotSad:2021}.
The details of this dimensional reduction in a constant uniform magnetic field with the fermions
in their lowest Landau level (LLL) are reviewed in Appendix~\ref{App:DimRed} for the convenience of the reader.

The energy density of chiral fermions in their LLL in $3+1$ spacetime dimensions $\ve(n_5)$ can be related
to the energy density of fermions in $1+1$ spacetime dimensions $\ve_2$ by
\be
\ve(n_5) = \frac{eB}{2\p} \, \ve_2(\tilde n)  = \frac{\p^2}{eB} \, n_5^2\,,
\ee
where
\be
n_5 = \frac{eB}{2\p}\, \tn
\label{eos2d}
\ee
is the $3+1$ dimensional chiral number density of fermions in the LLL, in terms of
\be
\tn = (-\tj^{\lsp a} \tj_a)^{\frac{1}{2}}\,,\quad a=0,1\,,\qquad\ve_2(\tilde n) =\sdfrac{\p}{2}  \tn^2
\label{tilden}
\ee
the chiral density and energy density of $1+1$ dimensional fermions restricted to moving in the
$\hat{\bf x}$ direction along $\vec B$. The 2D chiral current and number density in (\ref{tilden})
are denoted by $\tj_a$ and $\tn$ respectively.

Factoring out a common factor of the electron number density per unit area $eB/2\p$ in the LLL leads to
the 2D effective action
\be
S_{{\rm anom},B} = \frac{eB}{2\p}\! \int d^{\lsp 2} \bx_\perp \! \int d^{\lsp 2}x\,
\big[(\pa^a\h+ b^a)\tj_{a}+ \h\lsp{\mA}_2 - \ve_2 (\tn)\big]
\label{SfluidB}
\ee
of a 2D chiral superfluid, where the two dimensional axial anomaly is
\be
\mA_2 = \frac{e}{2\p}\, \e^{ab}F_{ab} = \frac{eE}{\p}\,,
\label{2Danom}
\ee
where the $a,b$ indices range over $0,1$ only and the electric field $E=F_{10}$ is also in the
$\hat{\bf x}$ direction along $\vec B$. The 2D axial chemical potential
\be
\tm = \frac{d\ve_2}{d\tn} = (-\pa^a\h\,\pa_a\h)^{\frac{1}{2}} = \p \tilde n = \m_5 = \frac{d\ve}{d n_5}
\label{mu2d}
\ee
is in fact the same as that evaluated in 4D. Note that in 2D the axial anomaly (\ref{2Danom}) is linear
in the electric field, so that variations in $\mA_2$ will be of the same order as those of $\h$ in
2D, or in 4D as a result of dimensional reduction in a classical background $\vec B$ field.

Apart from the overall factor of $\frac{eB}{2\p}\! \int\! d^{\lsp 2} \bx_\perp$, the 2D effective action
in (\ref{SfluidB}) for excitations along the $\vec B$ direction is thus~\cite{MotSad:2021}
\be
S_{\rm 2D} =\int d^{\lsp 2}x\, \big[(\pa^a\h+ b^a)\tj_{a}+ \h\lsp{\mA}_2
- \ve_2 (\tilde n)\big] = \int d^{\lsp 2}x\Big[- \sdfrac{\p}{2}\,
\pa^a \c\, \pa_a\c -\, (eE + \pa^a b_a)\, \c \Big]
\label{Seffchi}
\ee
in terms of a local pseudoscalar boson field $\c$, related to the chiral phase $\h$ by
\be
\h = -\p\c + \frac{\th}{2} \to -\p\c\,,
\ee
where the arbitrary constant $\th$ phase angle associated with the anomaly is set to zero here.
The 2D effective fluid action (\ref{Seffchi}) for excitations parallel to the $\vec B$ field
is then recognized as the bosonic effective action of the fermionic sector of the Schwinger model,
{\it i.e.}\ QED$_2$ of massless fermions~\cite{Schwinger:1962tp, Blaschke:2014,MotSad:2021}
in which the gauge potential in $\e^{ab}F_{ab}/2 = \e^{ab} \pa_a A_b$ of the 2D model
has been replaced by $e A_a + \e_a{\!}^{c} b_c $ of the dimensionally reduced 4D theory.

The variations
\be
\tj^{\lsp a} = \frac{\d}{\d b_a}\, S_{\rm 2D}[\c; A, b]  = \pa^a \c\,,\qquad
j^a= \frac{\d}{\d A_a}\, S_{\rm 2D}[\c; A, b] =-\e^{ab}\pa_b\c
\ee
express the bosonization rules for the 2D chiral and electric currents. The extremization
\be
\frac{\d}{\d \c}\, S_{\rm 2D}[\c; A, b]  = \p\,\sq_2\c - (E + \pa^a b_a)  =
\p \big(\pa^a \tj_{a} - \mA_2\big) = 0
\label{chi2}
\ee
recovers the 2D axial anomaly (\ref{anom2}), and the 2D wave operator $\sq_2$ shows that the $\c$ (or $\h$)
boson is a true propagating axion degree of freedom as a result of the anomaly itself. The Green's function of
the 2D wave operator $\left(\sq^{-1}_2\right)_{xy} = - \frac{1}{\p}\ln (x-y)^2$ in (\ref{SNL2}) is the massless
scalar propagator in $d=2$ spacetime dimensions, describing propagation along the $\vec B$ direction.
The non-local form of (\ref{Seffchi}) is
\begin{align}
	S_{\rm 2D}^{\text{NL}}[A, b] = -\frac{1}{2\p}\int
	d^{\lsp 2}x \int d^{\lsp 2}x'\,\big(\pa^a b_a+  E\big)_x  \left(\sq^{-1}_2\right)_{xx'}
	\big(\pa^a b_a+  E\big)_{x'}
\label{SNL2}
\end{align}
assuming the electric field ${\bf E} = E(t, x) \,\bf \hat x$ is also along the $\vec B$ direction and
independent of the transverse coordinates $\bf y$.

By the Gauss law $\pa_x E  = \r$, an electric field
\be
E  = -\frac{2\a B}{\p}\,\h
\label{Gauss2D}
\ee
parallel to the $\vec B$ field direction is induced. Finally, variation of
(\ref{SfluidB}) with respect
to $\h$ reproduces the axial anomaly back in $3+1$ dimensions
\be
\pa_\m J_5^\m= \frac{eB}{2\p^2} \big[\big(\pa_t^2 - \pa_x^2 \big) \h + \pa^a b_a\big]
= \frac{2 \a}{\p} \, EB = - \left(\frac{2 \a B}{\p}\right)^2 \h\,,
\label{strongBeoms}
\ee
which is equivalent to the linear wave equation
\be
\left(\pa_t^2-\pa_x^2 + \frac{2\a}{\p}\,eB\right)\h= -\pa^a b_a
\label{2Daxion}
\ee
with the mass gap $M^2 = 2 \a eB/\p$. The energy-momentum offset of the Weyl modes may be set
to zero, $b^a=0$, at this point. It has been retained in (\ref{2Daxion}) only to illustrate
that such an offset which is varying in time or space acts as a source for the axion phase
mode aligned with the magnetic field $\vec B$.

The wave equation (\ref{2Daxion}) describes chiral density waves (CDWs) and chiral magnetic waves
(CMWs) here propagating at the speed of sound $v_s= v_F =1$, with the quantum numbers of an axion.
The CDW/CMW $\h$ wave arises as a gapless bosonic mode of the axial anomaly describing excitations of
the one dimensional Fermi surface of chiral fermions in the LLL, and is a direct consequence of massless
anomaly $1/k^2$ pole. However, as a result of the dimensional reduction to 2D, the 2D anomaly (\ref{2Danom})
and $E$ in (\ref{Gauss2D}) is of the same order as $\h$ itself. Thus in this case the interaction
of $\h$ and $\mA_2$ cannot be neglected as in the 4D case of Sec.~\ref{Sec:ChiFlu4}, and as a result
the axion mode becomes massive (gapped) by its electromagnetic interactions, exactly as the Schwinger
boson does in the 2D Schwinger model, and analogously to the Anderson--Higgs mechanism in the Standard Model.
Thus by this dimensional reduction, a WSM subjected to a magnetic field gives rise to CDW/CMW axionic excitations
along the $\vec B$ direction, due to the axial anomaly itself, and furnishes a second example of ASB in
the absence of any direct fermion-fermion interactions, other than electromagnetic ones.

\section{Fermion Self-Interactions and SSB of Chiral Symmetry}
\label{Sec:NJL}

In the condensed matter literature on WSMs, it is usually supposed that the fermion excitations at the Weyl nodes may
be subject to a generic four-fermion interaction of the Nambu--Jona-Lasino (NJL) type~\cite{Wang:2012bgb,Nambu:1961tp,Nambu:1961fr}
\be
\sL_{\text{NJL}} = \tfrac{1}{4}\lsp G\lsp\big[(\bj\j)^2-(\bj\g^5\j)^2\big]
\label{NJL}
\ee
of unspecified origin in the material. This NJL Lagrangian is invariant under the global $U(1)^{\rm EM} \otimes U(1)^{\text{ch}}$ chiral symmetry $\j \to e^{i\a + i \b \g^5}\j$ just as the massless fermion Lagrangian
(\ref{Lagf}) is at the classical level. For $G >0$ the interaction (\ref{NJL})
is attractive. Since this fermion self-coupling $G$ has mass dimension $-2$, the NJL theory is non-renormalizable in the UV
and must come equipped with a cutoff at some high energy scale $\L$, corresponding to a short distance cutoff $1/\L$.

The fermion self-coupling $G$ is most conveniently handled by introducing a two-component charge neutral scalar field
\be
\F = \f_1 + i\f_2 \g^5\,,\qquad \F^{\dag} = \f_1 - i\f_2 \g^5
\label{Phidef}
\ee
in the Dirac matrix space, with both scalar ($\f_1$) and pseudoscalar ($\f_2$) components, coupled to the fermions
through the Yukawa interaction
\be
\mathscr{L}_Y = -g \bj\F\j = -g \bj\big(\f_1 + i \f_2 \g^5\big) \j\,,
\label{Yuk}
\ee
where $g$ is a dimensionless Yukawa coupling. Since the fermions transform under a
$U(1)^{\text{EM}} \otimes U(1)^{\text{ch}}$ phase transformation by (\ref{U1ch}), the interaction
(\ref{Yuk}) is invariant under this transformation provided
\be
\F \to e^{-2 i\b \g^5} \,\F = \big(\f_1 \cos 2 \b  + \f_2 \sin 2 \b \big)
+i \big(\f_2 \cos 2\b  - \f_1 \sin 2\b \big)\g^5\,,
\label{Phi-g5}
\ee
{\it i.e.}\ $\F$ is electrically neutral but carries a chiral charge that is opposite in sign
to that of $\j$ and twice as large. The interaction Lagrangian
\be
\sL_{\text{int}} = - g\bj\big(\f_1 + i \f_2 \g^5\big) \j - \frac{g^2}{G} \big(\f_1^2 + \f_2^2\big)
\label{Lint}
\ee
is equivalent to the original four-fermion NJL interaction Lagrangian (\ref{NJL}) after
a Hubbard--Stratonovich transformation, obtained by extremizing $\sL_{\text{int}}$ with respect
to $\f_i$, yielding
\be
g \f_1 = -\sdfrac{1}{2} G\lsp\bj\j \,,\qquad g \f_2 = -\sdfrac{i}{2} G\lsp\bj\g^5\j\,,\qquad \text{(classically)}
\label{scalars}
\ee
and substituting the result into $\sL_{\text{int}}$, whereupon $\sL_{\text{NJL}}$ is recovered.
Thus, the fermion bilinears of the NJL model are replaced by the $\F$ boson field in this description.

In the NJL model the $U(1)^{\text{ch}}$ symmetry is spontaneously broken, $\lag \bj\j\rag \neq 0$, and
the fermion acquires a mass gap $m$ if the self-coupling $G$ exceeds a certain critical value $G_c$. This
follows from the solution of the minimization of the effective potential for $\lag \bj\j\rag$, which leads to
condition
\be
2m \left( \frac{1}{G} - \frac{1}{G_c}\right) = - \frac{m^3}{4 \p^2} \ln\frac{\L^2}{m^2}\,,\qquad
G_c \equiv \frac{8\p^2}{\L^2}\,,
\label{NJLmin}
\ee
in terms of the ultraviolet cutoff $\L \gg m$. If $G < G_c$, the only solution of (\ref{NJLmin})
for real $m$ is $m=0$, the $U(1)^{\text{ch}}$ symmetry remains unbroken and the fermion remains
massless. On the other hand, if $G>G_c$ a second solution appears with $m>0$, given by
\be
\frac{m^2}{\L^2} \ln\frac{\L^2}{m^2} = 1 - \frac{G_c}{G} > 0\,,\qquad \text{for}\qquad G > G_c\,,
\label{SSBsoln}
\ee
in which the $U(1)^{\text{ch}}$ symmetry is spontaneously broken and the fermion acquires a mass gap $m> 0$. By
computing the second derivative of the scalar effective potential with respect to $\phi_1$, it is straightforward to show
that the symmetric solution at $m=0$ becomes unstable, {\it cf.}~(\ref{SecDeriv}), and the SSB solution (\ref{SSBsoln})
has a lower energy if $G> G_c$.  Thus, for sufficiently strong fermion self-interactions the fermions become gapped with
$m >0$, and the ground state breaks the $U(1)^{\text{EM}} \otimes U(1)^{\text{ch}}$ symmetry down to $U(1)^{\text{EM}} $.
As a result, Goldstone's theorem for the spontaneous breaking of a global symmetry assures us that there is a gapless
Goldstone boson corresponding to the chiral phase of the $\F$ field, in more than one spatial dimension.

These statements and the condition for SSB of $U(1)^{\text{ch}}$ have their analogs in the effective potential
of the bosonic field $\F$, once it is provided with a quartic self-interaction term $\l\, \lsp \tr\, \F^4$, so that
its energy is bounded from below. The $U(1)^{\rm EM} \otimes U(1)^{\text{ch}}$ invariant quartic potential
\begin{align}
V(\s) = \sdfrac{\k}{2}\,\s^2 + \sdfrac{\l}{4}\, \s^4
\label{Vpot}
\end{align}
depends only upon $\s = |\F|$ and the parameters $\k$ and $\l> 0$. The minimum of this potential at
\be
V'(\bar\s) = \k \bar\s + \l \bar\s^3 = 0\,,
\ee
which admits the solution
\be
\bs = \sqrt{\sdfrac{\!-\k\,}{\,\l}}\, \neq 0 \quad {\rm if} \quad \k< 0\,,
\label{barsig}
\ee
so that the $\F$ field develops an expectation value $\lag \f_1 \rag \equiv \bar \s \neq 0$ in the ground state
that spontaneously breaks the $U(1)^{\text{ch}}$ symmetry if $\k < 0$. Comparing the quadratic term in (\ref{Vpot})
to that of the fermion loop in (\ref{kaplamcalcEFT}) shows that
\be
\k=2 g^2 \Big(\frac{1}{G}-\frac{1}{G_c}\Big)
\label{kapNJL}
\ee
and the condition for SSB in the scalar theory $\k <0$ coincides with the condition $G>G_c$ in the
fermionic NJL model. Upon identifying also
\be
\l = \frac{g^4}{4\pi^2}\ln\frac{\L^2}{m^2}
\label{lamNJL}
\ee
from the quartic term induced by the fermion loop in (\ref{kaplamcalcEFT}), we then find that
\be
m^2 = g^2 \bs^2 = -g^2\, \frac{\k}{\l} = \frac{8 \p^2}{\ln (\L^2/m^2)}\, \Big(\frac{1}{G_c}-\frac{1}{G}\Big)>0
\qquad \text{for}\qquad G > G_c\,,
\label{massf}
\ee
which is equivalent to the gap equation (\ref{NJLmin}) in the fermionic NJL model with cutoff $\L$,
demonstrating the consistency of the scalar field description with the original NJL fermionic one.

Since the scalar $\F$ defined by (\ref{Yuk}) and (\ref{scalars}) has mass dimension one, the quartic potential
term  $\l\, \lsp\tr \F^4$ and kinetic terms for $\F$, Tr$(\pa^\m \F^\dag\lsp \pa_\m\F)$, are dimension four.
Although neither are generally considered in the condensed matter literature on WSMs, these terms are IR
relevant in the sense of the Wilson renormalization group and are necessarily generated also, with
coefficients logarithmically dependent upon $\L$, once the fermion one-loop quantum corrections to the NJL
model are considered, {\it cf.} Appendix \ref{App:Intoutfer}.

The classical $U(1)^{\rm EM} \otimes U(1)^{\text{ch}}$ symmetry (\ref{U1EM})-(\ref{U1ch}), prior to
consideration of the anomaly dictates that the kinetic terms for $\F$ must occur in the invariant combination
\be
\tr \big[\big(\pa^\m\F^\dag - 2i b^\m\g^5\F^\dag\big)\big(\pa_\m\F + 2i b_\m\g^5\F\big)\big]
\ee
and that the kinetic terms for the $A_\m$ and $b_\m = A^5_\m$ potentials involve only their respective field
strength tensors, namely $F_{\m\n}F^{\m\n}$ and $F^5_{\m\n}F^{5\,\m\n}$ should also be included for a UV
complete theory in general. A mixed $F_{\a\b}F^{5\,\a\b}$ term is disallowed by the discrete symmetries
of charge conjugation or parity spatial reflection.

These terms are generated by fermion loop integrations, with a logarithmic dependence upon the UV cutoff $\L$,
as shown in Appendix \ref{App:Intoutfer}, just as one would be expect by power counting and the $U(1)^{\rm EM} \otimes U(1)^{\text{ch}}$ classical symmetry. A UV renormalizable effective theory must contain these terms from the very beginning. In such a UV completion of the EFT
the couplings $\k, \l$ and $g$ and the fermion mass gap $m$ will all be be finite renormalized parameters independent
of the UV cutoff $\L$, to be fixed by experiment, and the scalar potential (\ref{Vpot}) can be treated at tree level,
as in the classical Landau theory of phase transitions~\cite{LandauLifStatPhys}.
This is the approach we take in the next section.

\section{A Renormalizable Theory for WSMs Encompassing both ASB and SSB}
\label{Sec:fullEFT}

The effective Lagrangian in the two cases of free or weakly interacting fermions and strongly self-interacting
fermions with SSB can be derived as different limits of one and the same UV renormalizable classical theory with Lagrangian
\begin{align}
\sL_{\text{cl}} = -\sdfrac{1}{4} F^{\mu\nu}F_{\mu\nu}
+ \bj\lsp\g^\m \big(i\lrpr_\m + eA_\m+b_\m\g^5\big)\j  - g\bj\F\j  + \sL_\F\,,
\label{Lcl}
\end{align}
with the bosonic Lagrangian
\begin{align}
\sL_\F &= -\sdfrac{1}{8}\, {\rm tr} \big[\big(\pa^\m\F^\dag - 2i\lsp b^\m\g^5\F^\dag\big)\big(\pa_\m\F + 2i\lsp b_\m\g^5\F\big)\big]
- \sdfrac{\k}{8}\, {\rm tr}\big(\F^\dag \F\big)
- \sdfrac{\l}{32}\,{\rm tr} \big(\F^\dag \F\big)^2 \nn
& = -\sdfrac{1}{2}\big(\pa_\m\f_1 - 2\lsp b_\m \f_2\big)^2
-\sdfrac{1}{2}\big(\pa_\m\f_2 + 2\lsp b_\m \f_1\big)^2
- \sdfrac{\k}{2}\, \big(\f_1^2 + \f_2^2\big)
- \sdfrac{\l}{4}\lsp \big(\f_1^2 + \f_2^2\big)^2\,,
\label{LPhi}
\end{align}
where we keep all possible relevant terms up to dimension four, fully invariant under the
$U(1)\otimes U(1)^{\text{ch}}$ and Lorentz symmetries. The couplings $g, \k, \l$ will now be arbitrary
renormalized ({\it i.e.}\ UV cutoff independent) parameters that ultimately would have to be fixed by experiment
in any given WSM. The fermions are taken to be massless initially, with any mass generated only through
the Yukawa interaction, and the scalar $\F$ developing a vacuum expectation value
$\lag \F\rag = \lag \f_1\rag \neq 0$.

As already discussed, one should also allow for kinetic terms for the axial vector potential, such as
$F_5^{\m\n}F_{5\lsp\m\n}/f_5^2$ for the transverse part of $b_\m$, with an independent normalization and
chiral coupling $f_5^2$ analogous to the electromagnetic coupling $e$ of  $\sL_{\text{cl}}$. Anticipating
that the $U(1)^{\text{ch}}$ symmetry will be broken by the axial anomaly, a kinetic term $(\pa^\m b_\m)^2$
for the longitudinal component with another independent coupling as well as a mass term $b^\m b_\m$ could also
be allowed in (\ref{Lcl}).  However, once the $U(1)^{\text{ch}}$ symmetry is explicitly broken by the axial
anomaly, there is nothing preventing this mass term from being large, of the order of the cutoff scale $\L$, in
which case all the components of the axial vector potential will be gapped and play no role in the low energy
EFT at long wavelength scales. Thus we omit from the effective Lagrangian $\sL_{\text{cl}}$ of (\ref{Lcl})
all such kinetic and mass terms involving the axial potential $b_\m$ from the outset.

The $U(1)^{\rm EM} \otimes U(1)^{\text{ch}}$ symmetry of the classical action
$S_{\text{cl}} = \int d^{\lsp 4} x \,\sL_{\text{cl}}$, with (\ref{Lcl}), (\ref{LPhi})
results in the electromagnetic and axial currents
\begin{align}
J^\m&= \frac{\d S_{\text{cl}}}{\d A_\m}=e\lsp\bj \g^\m \j\,,\label{Jdef}\\
J^\m_5 &= \frac{\d S_{\text{cl}}}{\d b_\m} = \bj \g^\m \g^5 \j +
4\,\f_2 \lrpr^{\llsp\m}\f_1  -4\lsp b^\m(\f_1^2 + \f_2^2)\,,
\label{J5def}
\end{align}
being conserved by Noether's theorem, independently of SSB or a non-zero fermion mass, upon use of the {\it classical}
equations of motion following from $S_{\text{cl}}$, {\it i.e.}\ before any consideration of the axial anomaly. Defining
the polar representation for $\F$,
\begin{align}
\F=\lsp\s \exp(2i \llsp\z\gamma^5)\,,
\label{Phipolar}
\end{align}
the bosonic part of the effective action (\ref{LPhi}) becomes
\begin{align}
\sL_\F &= -\tfrac{1}{2}\,\pa^\m\s\lsp\pa_\m\s
- 2\lsp \s^2 \lsp\big(\pa^\m \z+ b^\m\big)\big(\pa_\m \z+ b_\m\big) - V(\s)\,,
\label{Lboson}
\end{align}
and the axial current has both the fermionic and bosonic contributions
\begin{equation}
J^\m_5 = \bj \g^\m \g^5 \j - 4 \lsp\s^2 \lsp\big(\pa^\m\z+ b^\m\big)
\equiv J^\m_5[\j] + J^\m_5[\F]\,,
\label{J5fb}
\end{equation}
where
\begin{equation}
J^\m_5[\F] = 4\,\f_2 \lrpr^{\llsp\m}\f_1  -4\lsp b^\m(\f_1^2 + \f_2^2) =- 4 \lsp\s^2 \lsp\big(\pa^\m \z+ b^\m\big)
\label{J5Phi}
\end{equation}
is the bosonic part of the axial current.

Note that the bosonic terms in both (\ref{Lboson}) and (\ref{J5fb}) depend on the chiral phase $\z$ of (\ref{Phipolar})
only through the combination $\pa_\m \z + b_\m$, while the $U(1) \otimes U(1)^{\text{ch}}$ invariant quartic potential
(\ref{Vpot}) depends only upon $\s = |\F|$. In the presence of a non-zero $b^\m$, the chiral chemical potential is
identified as the invariant combination~\cite{Schmitt:2014eka}
\be
\m_5^2 = -(\pa^\m \z + b^\m)(\pa_\m \z + b_\m) = (\dot \z + b_0)^2 - (\na \z + \vec b)^2
\label{mu5z}
\ee
in the fluid description, analogous also to (\ref{mu5}), and thus the effective potential to be minimized with respect to
$\s$ when $\m_5 \neq 0$ is
\begin{align}
V_{\text{eff}}(\s, \m_5) = - 2\lsp\m_5^2\lsp \s^2 + V(\s)
\label{Veff}
\end{align}
showing that a non-zero chiral chemical potential enters $V_{\text{eff}}$ with a negative sign, always tending
to destabilize the $U(1)^{\text{ch}}$ symmetric state at $\s =0$. Indeed the condition
\begin{align}
V_{\text{eff}}'(\bs, \m_5)= - 4\lsp\m_5^2\lsp \bs + V'(\bs) = 0
\label{Veffmin}
\end{align}
with $\l>0$ admits the two possible solutions
\begin{subequations}
\begin{enumerate}[label=(\alph*),labelindent=3.5cm,leftmargin=*]
\item\label{casea} $\bs =0\,, \qquad  {\rm if} \quad 4 \lsp\m_5^2 < \k \,,\qquad U(1)^{\text{ch}}\ \rm symmetric\,,$
\hfill\refstepcounter{equation}\textup{(\theequation)}\label{SSBsiga}
\item\label{caseb} $\bs = \sqrt{\dfrac{4\lsp\m_5^2 -\k}{\l}}\neq 0 \,, \quad
{\rm if} \quad  4\lsp \m_5^2 >\k \,, \quad U(1)^{\text{ch}}\ \rm SSB.$
\hfill\refstepcounter{equation}\textup{(\theequation)}\label{SSBsigb}
\end{enumerate}
\label{SSBsig}
\end{subequations}
In the SSB case \ref{caseb}, (\ref{J5Phi}) together with (\ref{mu5z}) informs us that
\be
\frac{n_5}{\m_5} = \frac{1}{\m_5} \, \sqrt{-J^\m_5 J_{5\lsp \m}} = 4\lsp \bs^2
\label{identif}
\ee
connecting the EFT fluid description \ref{EFTiii} in the equilibrium ground state with the scalar field description.  By standard arguments~\cite{Schmitt:2014eka} this SSB state defines a relativistic superfluid with the polar angle
field $\z$ the Goldstone mode of $U(1)^{\text{ch}}$ SSB that dominates the long distance/low energy
spectrum and can be identified with the axion. Equation \eqref{identif} may be viewed as a definition of $n_5$, but the essential scale of SSB is of course $\bar{\sigma}$.

In the $U(1)^{\text{ch}}$ symmetric case \ref{casea} the bosonic fields $(\f_1, \f_2)$ are a gapped doublet with
equal mass squared $\k - 4\lsp\m_5^2 >0$. Thus they play no role at large distances and can be dropped
entirely in the low energy EFT, while with $g\bs = 0$ the fermions remain massless. In the limit of
non-self-interacting and massless fermions the axial anomaly must be taken into
account by adding the anomaly effective action (\ref{SNL4}) or (\ref{Sanom}) to the classical action of
(\ref{Lcl}). In other words in \ref{casea}, taking the quantum fermion loop and the axial anomaly into
account means making the replacement of the classical (\ref{Lagf}) coupled to electromagnetism by the
1PI effective action
\be
\mathscr{L}^{\text{\ref{casea}}}_{\text{eff}} = -\sdfrac{1}{4} F^{\mu\nu}F_{\mu\nu}
+ \bj\lsp\g^\m \big(i\lrpr_\m + eA_\m \big)\j + \big(\pa_\m\h + b_\m\big)\lsp J_5^\m + \h\lsp\mA_4
\label{Leffi}
\ee
in which the $\bj \g^\m\g^5 \j$ term of (\ref{Lagf}) is included in (\ref{Sanom}) and no fermion self-interactions
appear explicitly. Since the anomaly leads to the fermion pairing as in Fig.~\ref{Fig:epair}, single
fermonic excitations do not appear in the spectrum at zero temperature and the fermionic term in (\ref{Leffi})
can also be dropped, so that we obtain the low energy effective action in the ASB case
\be
\cS_{\text{ASB}} = \int d^{\lsp 4} x \left[-\sdfrac{1}{4} F^{\mu\nu}F_{\mu\nu}
+ \big(\pa_\m\h + b_\m\big)\lsp J_5^\m + \h\lsp\mA_4\right] = \cS_{\text{EM}}[A] + \cS_{\rm anom} [\h;A, b]
\label{Seffa}
\ee
in the $T=0$ vacuum, with $\cS_{\rm anom}$ given by (\ref{Sanom}). This is the low energy effective action
that applies for a WSM for Weyl nodes displaced in energy, when supplemented by the non-vacuum $-\ve(n_5)$,
as in Sec.~\ref{Sec:ChiFlu4}, or for a WSM placed in a strong magnetic field, with excitations aligned
with the $\vec B$ field in the dimensional reduction of Sec.~\ref{Sec:Bfield}. In both cases $\h$ describes
an axionic mode. The axion remains gapless if EM interactions can be treated as higher order, as in
Sec.~\ref{Sec:ChiFlu4}, or becomes gapped if they are of the same order, as in the effectively 2D case
of Sec.~\ref{Sec:Bfield}.

We shall next demonstrate that the low energy effective action in the SSB case \ref{caseb} leads to
essentially the same low energy effective action as (\ref{Seffa}), with the polar angle $\z$
of the scalar $\F$ field replacing the chiral angle and Clebsch potential $\h$ of $\cS_{\rm anom}$
in the ASB case \ref{casea}.

\section{Perturbations from Equilibrium: Superfluidity and the Goldstone Sound Mode}
\label{Sec:Perturb}

In terms of the parameters of the potential $V(\s)$, the equilibrium values of the energy density and pressure are
\begin{subequations}
\begin{align}
\bar \ve & = \k \bs^2 + \sdfrac{3}{4} \l \bs^4 \,,\\
\bar P & = \sdfrac{1}{4} \l \bs^4\,,
\end{align}
\label{epequil}\end{subequations}
where $\bs$ is given by (\ref{SSBsigb}). The absence of any axial current in equilibrium requires $\vec J_{\!5} = 0$ so that
\begin{align}
\na \bar \z = - \vec b\,,
\end{align}
while $\dot{\bar \z} =0$ and $\bar J^0_5= \bar n_5, \bar \m_5^2 = (b_0)^2$
in applying this EFT to WSMs in the case of Weyl nodes displaced in energy
and momentum as in Fig.~\ref{Fig:WeylCones}. From (\ref{identif}) and the minimization
condition (\ref{Veffmin}) at $\s = \bs$, we also have
\begin{align}
4 \lsp\bs^2\bar \m_5^2 = \bar \m_5\bar n_5 = \bs V'(\bs)
\label{mu5n5}
\end{align}
in the equilibrium ground state of $U(1)^{\text{ch}}$ SSB.

These equilibrium relations may also be used for small deviations away from equilibrium in the long wavelength limit of the EFT
in the effective boson description. Expressing the polar field variables as their equilibrium values plus small time and space
dependent perturbations, {\it i.e.}
\begin{subequations}
\begin{align}
\s &= \bs + \d \s \,,\\
\z &= \bar \z + \d \z\,,\\
\dot \z + b_0 &= \bar \m_5 + \d \dot \z\,,\\
\pa_i \z +  \vec b_i &= \pa_i(\d\z)
\end{align}
\label{perturb}\end{subequations}
with $b_0 = \m_5$, and expanding the action (\ref{Lboson}) to the second order in perturbations around the SSB ground state,
we find ({\it cf.}~\cite{Schmitt:2014eka})
\begin{align}
S^{(2)}_\F &= -\sdfrac{1}{2}\int d^{\lsp 4}x\, \big[\d\s(-\sq + M^2_\s)\d\s - 4\bs^2 \d\z \sq \d \z - 16 \bar \m_5\bs\,\d \s \, \d\dot\z\big]\,,
\label{quadAction}
\end{align}
where
\begin{align}
M^2_\s = V^{\prime\prime}_{\text{eff}}(\bs) = V^{\prime\prime}(\bs) -4\bar \m_5^2 = V^{\prime\prime}(\bs) - \frac{V'(\s)}{\bs} = 2\l \bs^2 =
2\,\big(4 \bar \m_5^2 - \k\big)\,.
\label{Msigma}
\end{align}
The last term of (\ref{quadAction}) shows that there is a mixing between the gapless
Goldstone mode $\d\z$ and gapped $\d\s$ mode, which affects the low energy Goldstone
mode. Substituting the complex Fourier decomposition
$\d \z \sim e^{-i\w t + i \vec k\cdot \vx}$ we find the $2 \times 2$ Hermitian matrix
\bea
\left(\begin{array}{cc}
-\w^2 + \vec k^2 + M^2_\s \quad & 4i \w  \bar \m_5\\
-4i \w \bar\m_5\quad & -\w^2 + \vec k^2
\end{array}\right)
\label{matrix}\eea
operating on the two-component vector $(\d\s, 2 \bs \d\z)$ of perturbations. Setting the determinant of this matrix to zero gives
a quadratic equation for $\w^2$, which yields the spectrum, consisting of one solution for $\w^2$ that is gapped at the scale
$M^2_\s$, and a second solution at
\be
\w^2 = v_s^2\, \vec k^2 + {\cal O}\! \left(\!\sdfrac{k^4}{M^2_\s}\!\right)\,,
\ee
which is a gapless acoustic Goldstone mode with speed of sound  that differs from the speed of `light' $c=1$
of the Weyl node Fermi velocity. This sound speed is given instead by
\begin{align}
v_s^2 &= \frac{M^2_\s}{M^2_\s + 16 \bar\m_5^2} =\frac{\bs \,V^{\prime\prime}(\bs) - V^{\prime}(\bs)}{\bs \,V^{\prime\prime}(\bs) +3 V^{\prime}(\bs)}
= \frac{\l\bs^2}{2 \k + 3\l\bs^2}
=\frac{d \bar P/d \bs}{d \bar \ve/d \bs} = \frac{d \bar P}{d \bar \ve}\,,
\label{spsound}
\end{align}
which agrees with that obtained from a hydrodynamic approach to a sound mode.

Thus although $v_s \neq 1$ due to the spontaneous breaking also of Lorentz symmetry by the condensate rest frame where $J^0_5 = n_5$ but $\vec J_{\!5} = 0$,  the Goldstone mode remains gapless. This is an axionic acoustic sound mode with speed (\ref{spsound}). Note also that this agrees with the $v_s^2 = 1/3$ and (\ref{epequil}) agrees with the equation of state $p = \ve/3$ of free massless fermions if and only if $\k = 0$,  in which case the EFT is conformal. Any $\k \neq 0$ corresponds to non-vanishing self-interactions of the fermions in the NJL description which breaks conformal invariance.

The eigenmode of (\ref{matrix}) propagating with the sound speed $v_s$ of (\ref{spsound}) is the linear combination
\begin{align}
&\big[k^2(1-v_s^2) + M_\s^2\big]\, \d \s - 8\bar \m_5 \bs\, \d \dot\z= 0 \nn
&\hspace{5mm}{\rm or}\quad\qquad \d\s \simeq \frac{8 \bar \m_5 \bs}{M_\s^2}\, \d \dot\z
\label{dsigdz}
\end{align}
to lowest order in $k^2/M^2_\s$ for long wavelength acoustic excitations of the superfluid. Thus for
this linear combination the second action to second order of the perturbations (\ref{quadAction}) becomes
\begin{align}
S^{(2)}_\F & \simeq \frac{1}{2} \left(\frac{4 \bs^2}{v_s^2}\right)
\int d^{\lsp 4} x\, \d\z \left(\!-\sdfrac{\pa^2}{\pa t^2} + v_s^2\, \na^2\right) \d\z
\nn & = \frac{1}{2} \frac{d n_5}{d \m_5}
\int d^{\lsp 4} x\, \d\z \left(\!-\sdfrac{\pa^2}{\pa t^2} + v_s^2\, \na^2\right) \d\z
\label{quadzeta}
\end{align}
for the Goldstone sound mode at long wavelengths. In obtaining this last relation (\ref{vels4}) and
(\ref{spsound}) has been used. Utilizing (\ref{J5Phi}), (\ref{perturb}) and (\ref{dsigdz}), the linear
perturbations in the axial current are
\begin{equation}
\d J^0_5 = 8 \bs b_0 \,\d \s + 4 \bs^2\, \d \dot \z = \frac{n_5}{\m_5}
\left(1 + \frac{16 \m_5^2}{M_\s^2}\right) \d \dot \z = \frac{n_5}{\m_5} \frac{1}{v_s^2} \, \d \dot \z
=\frac{d n_5}{d \m_5}\, \d \dot \z
\end{equation}
and
\begin{align}
\d \vec J_5 = -4 \bs^2 \na (\d \z) = - \frac{\bar n_5}{\bar \m_5} \na (\d \z)\,.
\end{align}
Thus, the axial anomaly equation becomes
\begin{align}
\pa_\m (\d J^\m_5) =
\frac{d n_5}{d \m_5}\left(\frac{\pa^2}{\pa t^2} - v_s^2 \na^2\right) \d\z
= \frac{2 \a}{\p\, } \d (\vec E \cdot \vec B)
\label{dJ5zeta}
\end{align}
if the anomaly source is also varied to linear order.

Comparing (\ref{dJ5zeta}) for the perturbations in the phase field $\d\z$ in the case of SSB
in the effective potential $V_{\rm eff}$ of (\ref{Veff}) with the perturbations in the
Clebsch potential $\d\h$ of the anomaly effective action (\ref{etaeom4}), we see that they
{\it coincide}, and with the same sound speed $v_s$ when $\k =0$, corresponding to the
case of free fermions considered in Sec.~\ref{Sec:ChiFlu4}. This shows the case of ASB
and more familiar SSB in fact describe the same low energy hydrodynamic effective theory
of a chiral superfluid in the sense of \ref{EFTiii} of the Introduction. This result, perhaps
at first sight surprising, may be understood from the following considerations.

In the SSB case \ref{caseb} the fermions acquire a mass gap $m=g\bs$ from the Yukawa interaction in (\ref{Lcl})
and no longer appear at distances greater than $1/m$. Thus they can be integrated out entirely at low
energies as in Appendix \ref{App:Intoutfer}. Since the EFT of (\ref{Lcl}) is renormalizable, the effect
of integrating out the fermions is to renormalize the parameters of the remaining bosonic effective theory,
and to add the finite anomaly action (\ref{etaonaxion}) in the bosonic sector. That the axial anomaly
survives intact even when the fermions become massive due to SSB is seen by the explicit computation of
the effective action, and in particular the term (\ref{etaonaxion}) of Appendix \ref{App:Intoutfer},
which coincides with the last term of (\ref{Sanom}) with $\h=\z$. Thus in this case \ref{caseb} of SSB the
effective bosonic action of the WSM becomes
\begin{equation}
\sL_{\rm eff}^{\text{\ref{caseb}}} = -\tfrac{1}{4} F^{\mu\nu}F_{\mu\nu} + \sL_\F  + \z \mA_4\,,
\label{Leffii}
\end{equation}
with $\sL_\F$ given by (\ref{Lboson}). However in $\sL_\F$ there remain both the angular phase mode $\z$ which
is gapless and the radial mode $\d \s$ with mass gap $M_\s$ of (\ref{quadAction})-(\ref{Msigma}).
The latter decouples at energy scales less than $M_\s$, so that the effective low energy action
for small fluctuations from equilibrium is only (\ref{quadzeta}), but this is exactly the same
result following from the fluid effective action (\ref{Sfluid}) in the ASB case for two Weyl nodes
displaced in energy, with $\z$ taking the place of $\h$. Thus in the SSB case \ref{caseb} the
effective action for low energy gapless excitations only is
\be
\cS_{\text{SSB}} = \int d^4 x \left[-\sdfrac{1}{4} F^{\mu\nu}F_{\mu\nu}
+ \big(\pa_\m\z + b_\m\big)\lsp J_5^\m + \z\lsp\mA_4 - \ve(n_5)\right]
\label{Seffb}
\ee
which is the same as (\ref{Seffa}) after the addition of the $-\ve(n_5)$ term and replacement of
$\h$ by $\z$. To lowest order in fluctuations from equilibrium the effective action in the SSB case
becomes
\be
\cS^{(2)}_{\text{SSB}} = \frac{1}{2} \frac{d n_5}{d \m_5}
\int d^{\lsp 4} x\, \d\z \left(\!-\sdfrac{\pa^2}{\pa t^2} + v_s^2\, \na^2\right) \d\z
+ \int d^{\lsp 4} x\, \d\z \mA_4
\label{zetafluc}
\ee
and its variation gives (\ref{dJ5zeta}), with $v_s^2$ given by (\ref{spsound}).

The reason for identification of the Clebsch potential $\h$ with the bosonic polar phase angle $\z$ of
the SSB phase is also not difficult to understand. Inspection of the Yukawa interaction in the polar representation
\be
\lsp \bj\lsp \F\lsp \j = \lsp\s\lsp \bj\lsp\exp\!\left(2i\z \g^5\right) \j
\label{Yukpolar}
\ee
shows that $\z$ of the scalar $\F$ can be shifted to the phase of the fermion fields by a $U(1)^{\text{ch}}$
transformation (\ref{U1ch}) by (\ref{Phi-g5}), with $\b = \z$. Then $e^{2i\z \g^5}$ is the complex phase
factor of the fermion condensate matrix $\lag \bj\j \rag$. The fact that $\h$ appears in the EFT
of ASB in (\ref{Seffa}) in exactly the same way as $\z$ does in the EFT of SSB in (\ref{Seffb})
implies that there must be a fermion condensate in the ASB with this chiral phase, induced
by the axial anomaly alone, even it is not immediately apparent from the introduction of $\h$
in (\ref{Sanom}). As discussed further in Sec.~\ref{Sec:Golds}, the cases of ASB and SSB, which
appear at first sight to be quite different, are both associated with formation of a fermion chiral
condensate which spontaneously breaks the $U(1)^{\text{ch}}$ symmetry, and leads to the same low energy
collective axion excitation, although by a different route.

\section{'t Hooft Anomaly Matching and Non-Decoupling of the Axial Anomaly}
\label{Sec:Hooft}

The effective action of the axial anomaly (\ref{Sanom}) is based on massless fermions, $m=0$.
As soon as the condition (\ref{SSBsigb}) for SSB is satisfied, $\bs \neq 0$, and the
fermion becomes massive, then for any finite fermion mass the fermionic contribution to the axial anomaly
$\cA$ is reduced according to (\ref{axdiv}). This $m$ dependent reduction in $\cA$ is the non-anomalous
contribution to the divergence of the chiral current $2i m \bj \g^5 \j$ for free massive Dirac fermions,
according to (\ref{Am}), and turns off the anomaly completely in the $m\to \infty$ limit, consistent with
decoupling of very heavy fermions in the Wilsonian effective action at low energies. This decoupling
occurs if the fermion mass is simply added to the free Dirac Lagrangian in (\ref{Lagf}),
explicitly breaking the $U(1)^{\rm ch}$ invariance of the massless theory.

\begin{figure}
    \centering
    \begin{tikzpicture}[decoration={snake, segment length=2.06mm, amplitude=0.7mm}]
    \tikzstyle{arr}=[decoration={markings,mark=at position 1 with
      {\arrow[scale=1.5]{>}}}, postaction={decorate}];
    \tikzstyle{arrmid}=[decoration={markings,mark=at position 0.5 with
      {\arrow[scale=1.5]{>}}}, postaction={decorate}]
    \draw[arrmid] (0,0)--(2,1.5);
    \draw[arrmid] (2,1.5)--(2,-1.5);
    \draw[arrmid] (2,-1.5)--(0,0);
    \draw[decorate] (3.5,1.5)--(2,1.5);
    \draw[decorate] (3.5,-1.5)--(2,-1.5);
    \draw[arr] (-1,0.25)--(-0.5,0.25) node[above, midway] {$k$};
    \draw[arr] (2.5,1.8)--(3,1.8) node[above, midway] {$p$};
    \draw[arr] (2.5,-1.8)--(3,-1.8) node[below, midway] {$q$};
    \filldraw[fill=black] (0,0) circle (1.5pt);
    \filldraw[cross,fill=white,thick] (-1.7,0) circle (4pt) node[left=3pt] {$J^\mu_5[\Phi]$};
    \draw[dashed] (-1.5,0)--(0,0) node[midway, below] {$\zeta$};
    \filldraw[fill=black] (2,1.5) circle (1.5pt) node[above=2pt] {$J^\alpha$};
    \filldraw[fill=black] (2,-1.5) circle (1.5pt) node[below=2pt] {$J^\beta$};
\end{tikzpicture}
    \caption{The triangle diagram that contributes to the bosonic part of the axial current vertex $J^\m_5[\F]$ of (\ref{J5Phi}) to two-photon amplitude $\G_5^{\m\a\b}(p,q)$, given by (\ref{bosonJ5}) in position space or (\ref{G5Phi}) in momentum space. Its contraction with $k_\m$ gives a contribution to the axial anomaly which
    is equal and opposite to the mass-dependent contribution of the fermion vertex $J_5^\m[\j]$ of (\ref{J5fb})
    given by (\ref{Am}).}
    \label{fig:zetaAAdiag}
\end{figure}
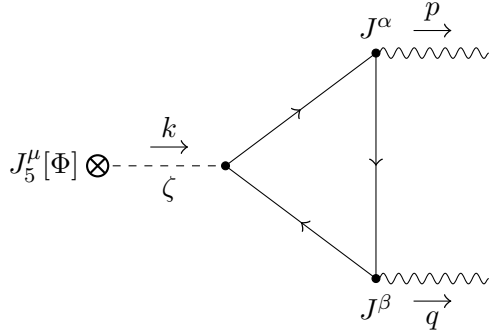

On the other hand if the fermion mass $m= g \bar \s$ is a result of SSB in the EFT of Sec.~\ref{Sec:fullEFT},
for which the axial current (\ref{J5def}) is a classically conserved Noether current, the mass suppression
of the anomaly in (\ref{Am}) should be canceled by the bosonic contribution to the axial current (\ref{J5Phi}).
That this is indeed the case can be shown from the classical equations of motion following from (\ref{Lcl}),
or in perturbation theory by computing the contribution to the axial anomaly of the bosonic term via its coupling
to the fermions (and hence electromagnetism) through the Yukawa interaction (\ref{Yuk}).
Expanding $ -g \lsp \bj\lsp \F\lsp \j$ to linear order in $\z$ in the polar representation gives the
contribution to the bosonic part of the axial current in (\ref {J5Phi})
\be
-4 \bs^2(-ig \bs)(2i) \int d^{\lsp 4}x'\, \big\lag \pa^\m\z(x)\z(x')\big\rag\,  \bj \g^5\j(x')
= -8g\bs^3 \pa_x^\m \int d^{\lsp 4} x'\, \big\lag \z(x) \z(x')\big\rag\,  \cP(x')
\label{bosonJ5}
\ee
in position space, where $\cP = \bj \g^5\j$. Using the definition of the canonically normalized Goldstone boson
propagator
\be
4 \bs^2 \big\lag \z(x) \z(x')\big\rag = -i \int \frac{d^{\lsp 4} k}{(2 \p)^4} \frac{e^{ik\cdot (x-x')}}{k^2}
\ee
and going over to momentum space, the bosonic contribution to the amplitude (\ref{trimom}) represented by
the diagram of Fig.\ \ref{fig:zetaAAdiag} is
\be
\G_{5\, \F}^{\m\a \b} (p,q) = -2\lsp m\, \frac{k^\m}{k^2}\, \L_5^{\a\b}(p,q)\,,
\label{G5Phi}
\ee
where $\L_5^{\a\b}(p,q) $ is given by (\ref{Am}) and $m= g\bs$ for the fermion mass has been used.
The contraction of (\ref{G5Phi}) with $k_\m$ therefore gives $-2\lsp m\lsp \L_5^{\a\b}(p,q)$ which is
equal and opposite in sign to the mass-dependent contribution of the fermion vertex $J_5^\m[\j]$ given
by (\ref{Am})~\cite{Bell:1969ts}.

Since the bosonic contribution to the non-anomalous divergence of the axial current (\ref{Jdef}) cancels the
fermionic mass contribution, the axial anomaly
\be
\pa_\m J^\m_5 = \mA_4 = \frac{e^2}{16\p^2}\, \e^{\a\b\m\n}F_{\a\b} F_{\m\n}
= \frac{2\a}{\p}\, \vec E \cdot \vec B\,,
\label{fullanom}
\ee
is completely unaffected by the onset of SSB and the fermion mass it generates. In fact this is a
necessary consequence of the Adler--Bardeen theorem which also holds in the $U(1)^{\rm EM}\otimes U(1)^{\rm ch}$
sigma model of Sec.~\ref{Sec:fullEFT}~\cite{AdlerBard:1969}.

Thus the fermions never decouple and continue to contribute the same amount (\ref{fullanom}) to the
axial anomaly which remains true {\it at any scale} in both the ASB case with nominally massless fermions and
the SSB case with the fermion mass generated by SSB of $U(1)^{\rm ch}$. This is an explicit demonstration
of 't Hooft anomaly matching from UV to IR scales, and its renormalization group invariance.  Even in the limit
$m\to \infty$ the full axial anomaly (\ref{fullanom}) remains unaltered and is simply transferred entirely
to the bosonic degrees of freedom, as shown by (\ref{anompi})-(\ref{actsofarI}) in Appendix \ref{App:Intoutfer}.

\section{The Goldstone Theorem for Anomalous Symmetry Breaking}
\label{Sec:Golds}

The Noether current (\ref{J5def}) of the classically $U(1)^{\rm EM} \otimes U(1)^{\rm ch}$ invariant
effective action $S_{\text{cl}} = \int d^{\lsp 4} x \,\sL_{\text{cl}}$, given by (\ref{Lcl}) and
(\ref{Lboson}), satisfies the identity
\be
\pa_\m J^\m_5 = i \,\frac{\!\d S_{\text{cl}}}{\d \j(x)\,} \g^5 \j + i \bj \g^5\frac{\d S_{\text{cl}}}{\d \bj(x)}
+ 2\lsp \f_2 \frac{\d S_{\text{cl}}}{\d \f_1(x)} - 2\lsp \f_1 \frac{\d S_{\text{cl}}}{\d \f_2(x)}\,,
\label{cldivJ5}
\ee
so that the axial current is classically conserved on-shell by use of the equations of motion following from
variation of the action $S_{\text{cl}}$. In the quantum theory this relation becomes a WT identity
\be
\left[\pa_\m \frac{\d }{\d b_\m(x)}  - i \,\frac{\d}{\d \j(x)}\, \g^5 \j - i \bj \g^5\frac{\d}{\d \bj(x)}
+ 2\lsp \f_1 \frac{\d }{\d \f_2(x)} - 2\lsp \f_2 \frac{\d }{\d \f_1(x)}\right]\G = \mA_4\,,
\label{anomWard}
\ee
for the 1PI effective action $\G[A,b,\j,\bj,\F]$ defined by the Legendre transform (\ref{Geffdef}),
(\ref{meanfield}), with the addition of the the anomalous axial current divergence
$\mA_4$ of (\ref{fullanom}) at right. By the non-decoupling and 't Hooft anomaly matching of
the previous section, the full anomaly (\ref{fullanom}) applies independently of the fermion mass
at all scales. If the fermions occur only in internal loops and no sources or mean fields are introduced
for them, the 1PI effective action $\G[A,b,\F]$ is independent of $\j$ and $\bj$, and the corresponding
terms in the anomalous axial WT identity may be dropped.

In the familiar situation of SSB with no anomaly, the WT identity (\ref{anomWard}) with $\mA_4=0$ is the basis
upon which Goldstone's theorem is established. In that case (\ref{anomWard}) is first integrated over the
spacetime volume $\int d^{\lsp 4} x$, selecting the Fourier component at zero four-momentum. Since the first
term on the left-hand side is a total divergence and the 1PI effective action $\G$ of (\ref{Geffdef}) by
definition does not contain one-particle singularities at $p^\m = 0$, this total divergence term gives
zero contribution. Then the result is functionally differentiated once more with respect to $\f_2(x')$ to obtain
\be
\int d^{\lsp 4} x \left[ \f_1 (x)\, \frac{\d^2\G }{\d \f_2(x)\d \f_2(x')}\right] = \frac{\d\G }{\d \f_1(x')}\,.
\label{WIf1f2}
\ee
If this is evaluated at the extremum of $\G$, at which the first variation at right vanishes, and if
this extremum occurs for a non-zero constant $\f_1 = \bs$, we obtain
\be
\bs \int d^{\lsp 4} x \left[\frac{\d^2\G}{\d \f_2(x)\d \f_2(x')}\right]
= - \bs\, G^{-1}_{22}(p)\big\vert_{p=0} = 0
\label{Golds}
\ee
in momentum space, since it follows from the definition of $\G$ in (\ref{Geffdef}) that the second variation
of $\G$ at left is (minus) the inverse propagator $G_{22}^{-1}$ of the $\f_2$ field. Since at zero four-momentum
$G_{22}^{-1}(0)$ is the $\f_2$ mass gap, (\ref{Golds}) shows that the $\f_2$ scalar is a massless Goldstone boson
in the SSB state where $\bs \neq 0$.

Now we may ask how this standard result is affected by the presence of the anomaly $\mA_4$ in the anomalous WT identity
(\ref{anomWard}). Note first that the anti-symmetric variation in (\ref{anomWard}) becomes
\be
\left[2 \lsp\f_1 \frac{\d }{\d \f_2(x)} - 2\lsp \f_2 \frac{\d }{\d \f_1(x)}\right] \G = \frac{\d \G}{\d \z(x)}
\ee
in the polar representation (\ref{Phipolar}). Thus repeating the step above of integrating over the spacetime volume
$\int d^{\lsp 4} x$, setting to zero the integral of the total divergence, but now functionally differentiating
with respect to the polar angle $\z(x')$ gives
\be
\int d^{\lsp 4} x\, \frac{\d^2\G }{\d \z(x)\d \z(x')} = \frac{ \d \mA_4 (x)}{\d \z(x')} = 0\,.
\label{WIzz}
\ee
Since the expression at left is proportional to the $\z$ field inverse propagator at $p^\m = 0$, the anomalous
term $\mA_4$ does not spoil Goldstone's theorem or give rise to a mass gap of the polar $\z$ mode,
provided that the anomaly $\mA_4$ itself is independent of $\z$, at least to first order at the extremum
of the effective action $\G$.

For the scalar EFT of Sec.~\ref{Sec:fullEFT}, (\ref{WIzz}) in the polar representation is clearly equivalent
to the standard form of (\ref{Golds}) by the change to polar field variables (\ref{Phipolar}). However,
the results of Secs.~\ref{Sec:ChiFlu4} and \ref{Sec:Perturb}, in particular comparison of (\ref{etaeom4})
and (\ref{dJ5zeta}) shows that they are identical also upon substitution of $\h$ for $\z$. One can check that
this general theorem holds in the EFT of Secs.~\ref{Sec:ChiFlu4} and \ref{Sec:Perturb},
with $\z$ is replaced by $\h$ in the former case. It fails in the 2D dimensional reduction of
Sec.~\ref{Sec:Bfield} because the 2D anomaly itself depends on the chiral rotation angle $\h=\z$
at linear order, by (\ref{2Danom}) and (\ref{Gauss2D}), so that the right side of (\ref{WIzz}) is
non-vanishing after dimensional reduction. This dependence generates an effective mass gap for the
would-be Goldstone mode for dynamical electric fields by the Anderson--Higgs mechanism.

Note that (\ref{Golds}) and (\ref{WIzz}) hold and the $\z$ angular mode is gapless even if Lorentz
invariance is broken and the $p^2$ terms in $G_{22}^{-1}(p^0, \vec p)$ or $G_\z^{-1}(p^0, \vec p)$ of (\ref{WIzz})
have unequal coefficients and the velocity of propagation differs from $c$ (or $v_F$). This is clearly
relevant to the case of Weyl nodes displaced in energy considered in Secs.~\ref{Sec:ChiFlu4} and
\ref{Sec:Perturb}, where a preferred rest frame is defined by the non-zero $\m_5$. In this case the
speed of propagation of the Goldstone mode is not $v_F$ (the surrogate for the relativistic speed of
light in a WSM), but the sound speed (\ref{vCDW}) or (\ref{spsound}) instead, determined by the equilibrium
equation of state and $dp/d\ve$. Although the CDW propagates at a different sound speed, it remains gapless,
and the Goldstone theorem (\ref{WIzz}) holds in the case of ASB as well as the more familiar SSB case,
resolving the apparent paradox of ASB giving rise to a Goldstone mode despite the anomaly explicitly
violating the $U(1)^{\text{ch}}$ symmetry. It also supports by explicit examples in WSMs the more abstract
argument that a gapless Goldstone mode should be expected also in the case of anomalous continuous symmetries,
based on the construction of non-invertible defect boundary operators~\cite{ChoiLamShao_PRL:2022,EtxeIqbal:2022}.

The alert reader will have noticed that (\ref{Golds}) is a relation for a canonically normalized
scalar field $\f_2$ and is multiplied by the expectation value $\bs$ in the SSB case. In the $U(1)^{\text{ch}}$
symmetric phase (\ref{SSBsiga}) $\bs = 0$ and (\ref{Golds}) becomes null: there is no Goldstone boson.
On the other hand (\ref{WIzz}) is expressed entirely in terms of the polar phase angle field $\z$, which is
dimensionless, and not a canonically normalized field (except in $d=2$). Referring to the explicit
examples of Secs.~\ref{Sec:ChiFlu4} or \ref{Sec:Perturb}, we see that the missing factor with the correct
dimensions is supplied by $dn_5/d\m_5$ in the case of Weyl nodes displaced in energy. Thus this quantity
proportional to $\bs^2$ must be non-vanishing for (\ref{WIzz}), or the corresponding relation
in terms of $\h$, and the $U(1)^{\text{ch}}$ symmetry must be spontaneously broken for (\ref{WIzz}) to
be non-null and a Goldstone boson to exist.

That $\bs$ must be non-zero for a Goldstone boson to exist leads to the conclusion that a fermion condensate
$\lag \bj \j\rag \neq 0$ must be present also in the ASB case of Weyl nodes separated by energy, at non-zero
$\m_5$, even though it does not appear explicitly in (\ref{etaeom4}) and (\ref{S2fluid}). This is also
indicated in the bosonic description by the fact that any non-zero chiral chemical potential with even free
massless fermions in Sec.~\ref{Sec:ChiFlu4}, corresponding to the critical value of $\k=0$ in $V(\s)$,
destabilizes the symmetric ground state to the broken symmetry state. This is the chiral analog of recent
studies of non-zero chemical potentials and charge densities being connected with SSB of $U(1)$ symmetry
and superfluid behavior as well~\cite{NicoPodoSant:2023}.

Since no direct four-fermion self-interaction was postulated in the ASB case for a system of apparently
`free' fermions, one may ask what the mechanism of symmetry breaking is that can produce a superfluid state
and Goldstone mode. Since the fermion pairing in the ASB case is due to the axial anomaly itself, one
must recognize that the Weyl fermions are not actually `free,' but still subject to the electromagnetic
interaction that leads to the anomaly. We are thus led to the conclusion that it is the same electromagnetic
interactions of the fermions responsible for the axial anomaly that are also responsible for the formation
of non-zero $\lag \bj \j\rag \neq 0$ condensate, spontaneously breaking the $U(1)^{\rm ch}$ symmetry in addition.

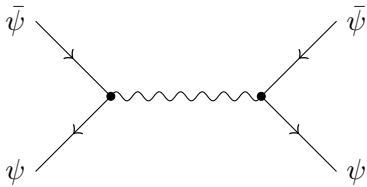
\begin{figure}[ht]
\centering
\begin{tikzpicture}[decoration={complete sines,
  segment length=1.5mm, amplitude=0.12cm, mirror, start up, end up}]
    \tikzstyle{arrmid}=[decoration={markings,mark=at position 0.5 with
      {\arrow[scale=1.5]{>}}}, postaction={decorate}]
    \draw[arrmid] (-1,1)--(0,0) node[at start, left] {$\bar{\psi}$};
    \draw[arrmid] (0,0)--(-1,-1) node[left] {$\psi$};
    \draw[arrmid] (3,1)--(2,0) node[at start, right] {$\bar{\psi}$};
    \draw[arrmid] (2,0)--(3,-1) node[right] {$\psi$};
    \draw[decorate] (0,0)--(2,0);
    \filldraw[fill=black] (0,0) circle (1.5pt);
    \filldraw[fill=black] (2,0) circle (1.5pt);
\end{tikzpicture}
\caption{The attractive channel of magnetic interactions of fermions leading to Cooper pairing.
The photon line must be dressed by interactions in the chiral plasma if Debye screening is present
at non-zero $\m_5$.}
\label{Fig:FourFer}
\end{figure}

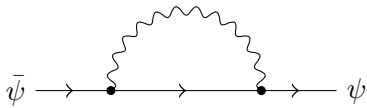
\begin{figure}[ht]
\centering
\begin{tikzpicture}[decoration={complete sines,
  segment length=1.5mm, amplitude=0.12cm, mirror, start up, end up}]
    \tikzstyle{arrmid}=[decoration={markings,mark=at position 0.5 with
      {\arrow[scale=1.5]{>}}}, postaction={decorate}]
    \draw[arrmid] node[left] {$\bar{\psi}$} (0,0)--(1,0);
    \draw[arrmid] (1,0)--(3,0);
    \draw[arrmid] (3,0)--(4,0) node[right] {$\psi$};
    \draw[decorate] (3,0) arc (0:190:1);
    \filldraw[fill=black] (1,0) circle (1.5pt);
    \filldraw[fill=black] (3,0) circle (1.5pt);
\end{tikzpicture}
\caption{The one-loop fermion self-energy $\S$ to be calculated for non-zero chiral potential $\m_5$ needed for
the gap equation in the ASB case. The photon propagator is the same as in Fig.~\ref{Fig:FourFer}. }
\label{Fig:Sigma}
\end{figure}

In fact it has been known for some time that long range magnetic interactions which are not Debye screened in
a plasma of finite charge density and chemical potential $\m$ lead to logarithmic divergences of the fermion
self-energy $\S \sim\ln (E-\m)$ for excitations $E$ close to the Fermi
surface~\cite{HolNorPin_PRB:1973,ChakNorSyl_PRL:1995}, an effect which is enhanced in degenerate relativistic
plasmas~\cite{Manuel:2000}, and which leads to non-Fermi liquid behavior of the plasma~\cite{GerIppRebhan:2004}.
The corresponding calculations have not been performed for a degenerate chiral plasma of massless fermions
at non-zero $\m_5$ to our knowledge. It is natural to conjecture however that the same unscreened long range
magnetic interactions provide an attractive channel for fermion pair interactions of Fig.~\ref{Fig:FourFer},
which generates an effective four-fermion interaction even if none were present initially. In that case
one would also expect that evaluation of the fermion self-energy in Fig.~\ref{Fig:Sigma} in the chiral
plasma will lead to a non-trivial solution to the fermion gap equation with a chiral condensate
$\lag \bj \j\rag \neq 0$ that spontaneously breaks $U(1)^{\text{ch}}$ symmetry in the case of non-zero
$\m_5$ as well.

If that is indeed the case, the superfluid axion CDW mode in the ASB case would be accounted for as the bound
state of the Cooper pairs produced by the anomaly itself as in Fig.~\ref{Fig:epair}, purely by their
electromagnetic interactions for arbitrarily small coupling $e$, consistent with the existence of the axial
anomaly due to those same interactions with electromagnetism. This same calculation at finite temperature $T$
and $\m_5$ should also determine the critical temperature $T_c\sim \m_5$ at which the condensate first vanishes
and with it the Goldstone mode and superfluid behavior. The fermion anomaly diagram of Fig.~\ref{Fig:J5JJ}
itself should also be calculated at non-zero $\m_5$ to directly verify the existence of the gapless pole
with the CDW sound speed  $v_s^2 = \frac{1}{3}v_F^2$ of (\ref{vCDW}), and complete the picture in the ASB case.

\section{Summary and Discussion}
\label{Sec:Sum}

Exploiting the formal identity of gapless Weyl nodes in WSMs to massless Weyl fermions in relativistic QFT,
we have proposed in this paper a UV renormalizable effective theory in Sec.~\ref{Sec:fullEFT} that
encompasses both the case of SSB by four-fermion interactions of the NJL kind as in Sec.~\ref{Sec:NJL}, and
the case when these interactions between the Weyl nodes are weak and below the SSB threshold of
$G \le G_c$ or $\k \le 0$. This provides a consistent QFT framework and basis for deriving controlled
low energy EFT limits for WSMs in both cases.  There is of course no claim that (\ref{Lcl}) represents
the `true' microscopic degrees of freedom of a WSM, which depend upon the electronic valence and conduction
bands of the material. The point is rather that by expressing the EFT in terms of a small number
of finite parameters that are insensitive to short distance physics, which can be determined
by measurements in each WSM, the Landau paradigm of EFTs of types \ref{EFTi} and \ref{EFTiii} can be
found for WSMs, and their consequences reliably deduced independently of the microscopic physics.
Integrating out the Weyl fermions when they become gapped can also be studied, as in
EFTs of type \ref{EFTii} in Appendix \ref{App:Intoutfer}.

That the axial anomaly is a consequence purely of the symmetries of the ground state, independently of
high energy or short distance physics has been emphasized in Appendix \ref{App:Axialanom}, to elucidate
why it plays an important role in the low energy EFT of a WSM. The massless $1/k^2$ pole and finite
sum rule associated with the anomaly are low energy features of the axial anomaly, which leads to a propagating
collective axion mode that should be present in WSMs even for cases in which direct fermion-fermion
interactions are weak or absent. In the framework proposed by starting with (\ref{Lcl}),
the axial anomaly and the axionic collective mode to which it leads can be derived from first principles of QFT
in WSMs, which directly gives a superfluid EFT of type \ref{EFTiii}. This axion excitation is
a collective mode of correlated fermion pairs at the Fermi surface, and arises in a zero temperature
WSM, whether or not there are strong four-fermion interactions of the Weyl modes. In the case
of strong four-fermion interactions the axion is a Goldstone mode of the SSB of $U(1)^{\rm ch}$
symmetry breaking with a non-zero scalar order parameter $\bs$ in its ground state.

Eq.~(\ref{WIzz}) shows that despite the axial anomaly explicitly contributing an anomalous term to the
$U(1)^{\rm ch}$ WT identity in (\ref{anomWard}), it does not spoil Goldstone's theorem as long as $\mA_4$
itself is independent of $U(1)^{\rm ch}$ rotations to first order, when evaluated at the ground state
extremum of the effective action $\G$. Thus the effective action of the axial anomaly itself
(\ref{Sanom}) predicts the existence of a propagating gapless CDW which is linearly coupled to the
$\vec E \cdot \vec B$ anomaly, which can be identified as an axion even in WSMs with weakly interacting
Weyl modes, where such a collective boson mode might not have been expected. This extends Goldstone's theorem
to the ASB case, supporting more formal arguments in \cite{ChoiLamShao_PRL:2022,EtxeIqbal:2022} with an
example of laboratory realizable system of anomalous symmetry breaking.

In both the SSB and ASB cases the axion is a collective Goldstone excitation of a low
temperature superfluid phase of a WSM, as in both cases the low energy EFT is of the
fluid form \ref{EFTiii} of a relativistic superfluid with the Fermi velocity $v_F$ of the
Weyl nodes replacing the speed of light $c$ of relativistic QFT. In both cases the low-energy
EFT of axions $\h=\z$ in WSMs takes the form
\be
S_{\rm WSM} = \int d^{\lsp 4} x \left[-\sdfrac{1}{4} F^{\m\n}F_{\m\n}
+ \big(\pa_\m\h + b_\m\big) J^\m_5 - \ve(n_5)+ \h\lsp\mA_4\right]
\label{SWSM}
\ee
where $b^\m = (b^0, b^i)$ is the axial vector potential, whose equilibrium value is one-half the separation of
Weyl nodes in energy and spatial momentum respectively, {\it cf.} Fig.~\ref{Fig:WeylCones}, and $\mA_4$
is the axial anomaly (\ref{divJ5}) which is independent of scale. In the case of Weyl nodes displaced
from each other in energy, $\m_5 \neq 0$, {\it cf.} (\ref{identif}), the gapless axionic excitation from
equilibrium following from (\ref{SWSM}) is described by the quadratic action (\ref{zetafluc}) and the linear
equation of motion (\ref{dJ5zeta}) with $\d \z = \h$, together with Maxwell's equations following from
(\ref{SWSM}). For a WSM placed in a strong enough magnetic field such that the fermions are all in their
lowest Landau level, (\ref{SWSM}) reduces instead to the two-dimensional effective action (\ref{SfluidB})
or (\ref{Seffchi}), for excitations along the $\vec B$ direction. This dimensionally reduced effective
action is equivalent to massless QED$_2$, {\it i.e.}\ the Schwinger model~\cite{Schwinger:1962tp}, in which
case the axionic excitation acquires the squared mass gap $M^2 = 2 \a eB/\p$, in analogy with the
Anderson--Higgs mechanism in a superconductor.

The $\h$ variable was introduced in (\ref{Sanom}) as a Clebsch potential in an effective fluid
description, with no apparent scalar order parameter of spontaneous $U(1)^{\rm ch}$ symmetry breaking
to which it is attached. Yet the fact that the axionic mode in this case obeys exactly the same
equation of motion (\ref{etaeom4}) as the phase mode $\z$ (\ref{dJ5zeta}) of the bosonic order parameter
is a clear indication that the two cases of ASB and SSB are in fact very closely related.
In particular a chiral condensate $\lag \bj \j\rag$ scalar order parameter that spontaneously
breaks the $U(1)^{\rm ch}$ symmetry (even with the anomaly present) should exist for which
$\exp(2 i\h)$ is its complex phase. Verifying this will require finding a non-trivial solution
of the gap equation resulting from the self-energy of Fig.~\ref{Fig:Sigma} with $\lag \bj \j\rag \neq 0$.
In that case the spontaneous breaking of the classical $U(1)^{\rm ch}$ symmetry would be due
to fermion pairing through the attractive channel of Fig.~\ref{Fig:FourFer} and the axion mode
of Fig.~\ref{Fig:epair} the result of the same electromagnetic interactions that give rise
to the axial anomaly itself. This same self-energy at finite temperature $T$ should determine when
the coherent state Cooper pairing dissociates and the superfluid to normal Fermi liquid fluid phase
transition takes place. This would be expected to be at temperatures $T\sim \m_5 = (\m_5/10 {\text{ meV}})\, 116$ K.

The WSM realization of axionic excitations opens new opportunities for experimental tests of the theoretical ideas presented in this paper. For example, the frequency dependence of the AC current in \eqref{Janom} induced in external electromagnetic fields with non-zero $\vec{E}\cdot\vec{B}$ may be used to experimentally probe the propagating axionic mode. The axionic mode and resulting current response may also be induced by other sources, e.g.\ external strain fields. We will discuss these and other potential experimental signatures in a forthcoming publication.

The EFT approach to WSMs presented here provides an interesting application of the 't Hooft anomaly matching condition
across scales from the UV renormalizable (\ref{Lcl})-(\ref{LPhi}) to the macroscopic superfluid effective
theory of emergent axion excitations of (\ref{Seffa}) or (\ref{Seffb}). An intriguing possibility is that
additional insights into the operation of the axial anomaly, anomaly matching and possible emergent axionic
mode in the Standard Model of particle physics could result from detailed study of the properties and
excitations of WSMs, to which the EFT of this paper can be applied.

\ack{The authors would like to thank M.\ Chernodub, N.\ Iqbal, K.\ Landsteiner, C.\ Reichhardt, A.\ Saxena, and M.\ Vozmediano for comments on this work. The work of AVS is partly supported by the European Research Council project ERC-2018-ADG-835105 YoctoLHC, by the Marie Sklodowska-Curie Individual Fellowship under JetT project (project reference 101032858),  and by Funda\symbol{"00E7}\symbol{"00E3}o para a Ci\symbol{"00EA}ncia e a Tecnologia (FCT) under contract 2022.06565.CEECIND. AVS also acknowledges further support from the Maria de Maetzu excellence program under projects CEX2020-001035-M; from the Spanish Research State Agency under project PID2020-119632GB-I00; from Xunta de Galicia (Centro singular de investigación de Galicia accreditation 2019-2022), and from European Union ERDF. AS is supported by the Royal Society under grant URF{\textbackslash}R1{\textbackslash}211417 and by STFC under grant ST/X000753/1.}

\begin{appendices}

\section{The Chiral Anomaly in \texorpdfstring{QED$_\mathbf{4}$}{QED\_4} and its IR Aspect}
\label{App:Axialanom}

In this appendix we briefly review the evaluation of the fermion triangle amplitude in Fig.~\ref{Fig:J5JJ},
\be
\G_5^{\m\a\b}(p,q) = i \int d^{\lsp 4} x \int d^{\lsp 4} y\, e^{i p\cdot x + i q \cdot y}\
\lag 0| {\cal T} J_5^{\m}(0) J^{\a}(x) J^{\b}(y)|0\rag\Big\vert_{A=0}
\label{trimom}
\ee
in momentum space by imposing the appropriate symmetries of the low energy theory.
Conventions used in this and subsequent appendices for Dirac $\g$-matrices satisfying the anti-commutation property $\{\g^\m,\g^\n\}= -2\,\h^{\m\n}$ with $\eta_{\mu\nu} = {\rm diag} (-1,1,1,1)$, are such that $\g^0 = (\g^0)^\dag$ is Hermitian while the $\g^i = -(\g^i)^\dag$ are anti-Hermitian.
The chirality matrix $\g^5 \equiv i \e_{\a\b\m\n}\g^\a\g^\b \g^\m \g^\n/4! = i \g^0\g^1\g^2\g^3 = (\g^5)^\dag$
(with $\e_{0123}=1$) is also Hermitian, with eigenvalues $\pm 1$ corresponding to right-handed or
left-handed chiralities respectively.

First, observe that a possible linear UV divergence of $\G_5^{\m\a\b}(p,q)$ proportional to an
arbitrary four-vector, if present, would imply that the vacuum or ground state
is not Lorentz invariant. Hence the first requirement \ref{inva} of Lorentz invariance automatically forces this possible linear divergence to zero identically. Explicitly, since
$\G_5^{\m\a\b}(p,q)$ must be a function only of the two independent four-momenta $p$ and $q$,
and with three Lorentz indices must transform as a third-rank tensor, it follows
that $\G_5^{\m\a\b}(p,q)$ must be expressible as a sum of all possible third-rank
tensors $\t_i^{\m\a\b}(p,q)$, each of which can depend only upon the
two independent external $4$-momenta $p^\m$ and $q^\n$. If the vacuum ground state is also invariant under the discrete symmetry
of spatial reflection parity (P) combined with time reversal (T), then each of the $\t_i^{\m\a\b}(p,q)$ basis tensors must also involve one factor
the totally antisymmetric Levi--Civita pseudo-tensor $\e^{\a\b\r\s}$, since $J^\m_5$ and hence $\G_5^{\m\a\b}$ is odd under P and T separately, but invariant (even) under the combined operation of PT.

Next, electric current conservation at the two electromagnetic current vertices $J^\a$ and $J^\b$ requires that each of the basis tensors $\t_i^{\m\a\b}(p,q)$ into which (\ref{trimom}) can be expanded must satisfy
\be
p_\a\, \t_i^{\m\a\b}(p,q)=0=q_\b\,\t_i^{\m\a\b}  \,, \qquad {\rm for\ each}\ i\,.
\label{taucons}
\ee
Defining the psuedo-tensor $\y^{\a\b}(p,q)$ by (\ref{upsdef}), which satisfies
\be
p_{\a} \y^{\a\b}(p,q) = 0 = q_{\b}\y^{\a\b}(p,q)
\ee
it is easily shown by exhaustive enumeration that there are exactly six tensors meeting all of the symmetry
requirements of $\t_i^{\m\a\b}(p,q)$, which are listed in Table \ref{chitens}~\cite{GiaEM:2009}.

\begin{table}[H]
$$
\begin{array}{|@{\hspace{.8cm}}c @{\hspace{.8cm}}|@{\hspace{.8cm}}c @{\hspace{.8cm}}|}
\hline
i & \t_i^{\m\a\b}(p,q) \\ \hline \hline
1 & -\pq \,\e^{\m\a\b\g} p_\g  -
p^\b \,\y^{\m\a}(p,q) \\ \hline
2 & p^2 \e^{\m\a\b\g} q_\g \, +
 p^{\a} \y^{\m\b}(p,q) \\ \hline
3 & p^\m\,\y^{\a\b}(p,q)\\ \hline \hline
4 & \pq \,\e^{\m\a\b\g} q_\g  +
q^{\a} \,\y^{\m\b}(p,q) \\ \hline
5 & -q^2 \e^{\m\a\b\g} p_\g \,
-q^{\b} \y^{\m\a}(p,q)\\ \hline
6 & q^\m\,\y^{\a\b}(p,q) \\ \hline
\end{array}
$$
\caption{The $6$ third-rank Lorentz pseudo-tensors linear in  $\e^{\a\b\r\s}$ satisfying (\ref{taucons}),
only 4 of which are linearly independent due to the Schouten identity~\cite{GiaEM:2009,ArmCorDRosaPRD10}. }
\label{chitens}
\end{table}

Thus, general conservation and invariance principles prescribe that the triangle amplitude (\ref{trimom}) must be expressible as
\be
\G_5^{\m\a\b}(p,q) =  \sum_{i=1}^6 f_i(k^2; p^2, q^2; m^2) \, \t_i^{\m\a\b}(p,q)\,,
\label{Amp5mom}
\ee
where the form factors $f_i = f_i(k^2; p^2,q^2)$ are Lorentz scalar functions of the three independent
Lorentz invariants $p^2,q^2$ and $k^2 = (p+q)^2$. Since each of the six basis tensors satisfying (\ref{taucons})
in Table \ref{chitens} are homogeneous of degree $3$ in the external momenta, $3$ powers of momenta have been
extracted from the triangle amplitude $\G_5^{\m\a\b}(p,q)$. Hence the remaining scalar amplitude functions $f_i$ are of degree $-2$ therefore may be expressed
as Feynman integrals that are quadratically {\it convergent}, {\it i.e.}\ independent of any high energy
or short distance cutoff. The result is that requirements (\ref{inva}) and (\ref{invb})
render the triangle amplitude $\G_5^{\m\a\b}(p,q)$ UV finite and well-defined.

The full amplitude (\ref{Amp5mom}) must also be Bose symmetric,
\be
\G_5^{\m\a\b}(p,q)=\G_5^{\m \b \a}(q,p)
\label{chibose}
\ee
under simultaneous interchange of $p, q$ and $\a,\b$. Owing to the Bose symmetry and overcompleteness of the six tensors in Table \ref{chitens}, finally only
two of the scalar coefficient functions, $f_1$ and $f_2$ say, need to be independently computed.
The finite scalar coefficient functions are given in the literature \cite{Rosenberg:1963, Adler:1969gk},
and most conveniently expressed in terms of integrals
\be
f_i(k^2; p^2, q^2; m^2) = \frac{e^2}{\pi^2}\int_0^1 dx\int_0^{1-x} dy\ \frac{c_i(x,y)}{D}
\label{fici}
\ee
over the Feynman parameters $(x,y)$ with $0\le x+y\le 1$, where the $c_i(x,y)$ are the simple polynomials
\be
\begin{aligned}
c_1(x,y) &= c_4 (x,y) = xy\,,\\
c_2(x,y) & = c_5(y,x) = x(1-x)\\
c_3(x,y) &= c_6(x,y) =0\,,
\end{aligned}
\label{ci}
\ee
and the denominator $D$ of (\ref{fici}) is
\begin{align}
D &= p^2 x(1-x) + q^2 y(1-y) + 2\pq \, xy + m^2 \nn
& = (p^2\,x +q^2\,y)(1-x-y)+ xy\,k^2+m^2 \equiv xy \big (k^2 + S\big)\,,
\label{denom}
\end{align}
which defines the quantity $S(x,y; p^2, q^2; m^2)$ for arbitrary finite fermion mass $m$. Both $D$ and $S$
are strictly positive for spacelike momenta, $k^2, p^2, q^2 > 0$.

Computing the contraction with the momentum $k_{\m} = (p + q)_{\m}$ entering at the axial vector vertex results in the
well-defined finite result (\ref{axanom}) with
\begin{align}
{\cA}(k^2; p^2,q^2;m^2)  &= 2\pq\, f_1 + p^2 f_2 + q^2 f_5
=\frac{e^2}{\p^2}\int_0^1 dx\int_0^{1-x} dy\  \frac{D- m^2}{D}\nn
&=\frac{e^2}{2\p^2}\left(1 - 2m^2\,\int_0^1 dx\int_0^{1-x} dy\,\frac{1}{D}\right)
\label{axdivA}
\end{align}
by (\ref{fici}), (\ref{denom}) and Table \ref{chitens}, which is (\ref{axdiv}) of Sec.~\ref{Sec:TriAnom}.
The second term proportional to $m^2$ is what would be expected from the axial vector divergence $\pa_\m J_5^\m
= 2im \bj\g^5\j$, {\it cf.} (\ref{divJ}), following from use of the on-shell Dirac equation for fermions
of mass $m$, {\it i.e.}
\be
2 im\! \int\! d^{\lsp 4} x \! \int\! d^{\lsp 4}y\,  e^{ip\cdot x + i q\cdot y} \big\lag \cP(0) J^\a(x) J^\b(y) \big\rag \Big\vert_{A=0}
\!\equiv 2\lsp m\lsp\L_5^{\a\b}(p,q) = -\frac{e^2m^2}{\p^2}\, \y^{\a\b}(p,q)\! \int_0^1\! dx \!\int_0^{1-x}\!\!dy\, \frac{1}{D}
\label{Am}
\ee
in the same Feynman parameter representation, with $\cP=\bj\g^5\j$ the pseudoscalar density \cite{Horejsi:1985}.

If the fermion mass $m=0$, this last term of (\ref{axdivA}) proportional to $m^2$ vanishes, and (\ref{A0}) is the
finite and non-zero axial current anomaly. On the other hand, in the limit $m^2\to \infty$, the denominator $D$
of (\ref{denom}) can be replaced by $m^2$, and the entire $\cA$ of (\ref{axdivA}) is canceled since
$2\int_0^1 dx \int_0^{1-x} dy = 1$, hence $\cA = 0$, so that the fermion decouples entirely and
there is no anomaly pole in this limit.

One should recognize that conservation at the axial vector vertex $J_5^\m$ could be required, by defining
$\G_5^{\m\a\b}(p,q)$ to violate the charge conservation condition \ref{reqii} instead. The anomaly is
fundamentally a conflict between symmetries, forcing a choice between the classical conservation
equations (\ref{divJ}), which can only be decided by additional physical input. Since charge conservation
is well-established, and it is known that the $m \to 0$ limit is not equivalent to the $m=0$ theory
when $e \neq 0$, which limit is beset with {\it infrared} (IR) divergences, {\it cf.}~\cite{LeeNau:1964,DolZak:1971,Huang}
and Appendix \ref{App:Axialanom}, the choice is made to preserve $U(1)^{\text{EM}}$ and discard $U(1)^{\text{ch}}$
as a fundamental symmetry.

This derivation of the chiral anomaly and anomaly pole relies only upon the low energy symmetries of the theory. Thus exactly the same anomaly is obtained by any method
which satisfies these same symmetry requirements, among them Pauli--Villars regularization~\cite{Horejsi:1992}, dimensional
regularization~\cite{Bertlmann:2000}, heat kernel methods~\cite{Vassil:2003}, or Fujikawa's regularization of the fermion
functional integral for the axial current~\cite{Fujikawa:1979ay, Fujikawa:1980eg}. The fact that these various methods were
developed in QFT to regularize its short distance behavior, as the first step to removing UV divergences in a renormalization
procedure, led to some obscuring of the physical basis of the axial anomaly as an essentially IR phenomenon.
The only role these more sophisticated regularization methods play is simply to require $\G_5^{\m\a\b}$ to satisfy conditions
\ref{inva}-\ref{invd}, which are properties of the vacuum or ground state of the theory.

To emphasize its independence of extreme high energy or short distance physics, and relation to fermion pairing,
one may also derive (\ref{axanom}), (\ref{axdiv}) by a dispersion approach. For this one observes that the absorptive
part of the triangle amplitude (\ref{Amp5mom}) is determined by the on-shell matrix elements $\sum_n\lag 0 |J_5^\m|n\rag \lag n| J^\a J^\b|0 \rag$ for timelike $k_\m$, with $\vert n\rag$ a complete set of two-particle intermediate fermion/anti-fermion
states. This on-shell absorptive part is well-defined and finite, and in no need of regularization. Then the full amplitude
is obtained from the Kramers--Kronig dispersion relation
\bes
\bea
&&f_i(k^2;p^2,q^2;m^2) =\int_0^\infty \frac{ds}{k^2 + s}\ \r_i (s;p^2,q^2;m^2)\,,
\label{fDisp}\\
&&{\rm Im}\big(f_i(k^2 \!-\! i\e;\,p^2,q^2;m^2)\big)= \p  \r_i (s ;p^2,q^2;m^2)\Big\vert_{s=-k^2}
\eea
\label{KraKro}\ees
in the complex $k^2$ plane. Inserting $1 = \int_0^\infty ds\ \d (s-S)$ into (\ref{fici}), and using $D=xy(k^2 + S)$
from (\ref{denom}) allows us to express these absorptive parts by
\be
\r_i(s; p^2,q^2;m^2) = \frac{e^2}{\p^2} \int_0^1 dx \int_0^{1-x} dy \ \frac{c_i(x,y)}{xy}
\ \d\big(s - S(x,y;p^2, q^2; m^2)\big)
\label{ri}
\ee
in the Feynman parameter integral representation. Using (\ref{ci}) it follows from (\ref{ri}) that the combination
of absorptive parts appearing in (\ref{axdiv}) or (\ref{axdivA})
\be
2\pq\, \r_1 + p^2 \r_2 + q^2 \r_5 =  (k^2 + s) \, \r_1
- \frac{e^2 \,m^2}{\p^2} \int_0^1\! dx\int_0^{1-x} \!\frac{dy}{xy} \ \d\big(s - S(x,y;p^2, q^2; m^2)\big)
\label{sumrhoi}
\ee
which vanishes at $s=-k^2$ when $m=0$. Thus there is no anomaly in the absorptive part of (\ref{Amp5mom}).

On the other hand from $c_1(x,y)=xy$ given in (\ref{ci}), we have from (\ref{ri}) that
\be
\r_1(s; p^2,q^2;m^2)  = \frac{e^2}{\p^2} \int_0^1 dx \int_0^{1-x} dy \ \d\!
\left(\!s - \frac{(p^2x + q^2y)(1-x-y) + m^2}{xy}\right)\neq 0\,,
\label{rho1}
\ee
so that using this in (\ref{axdivA}) with (\ref{sumrhoi}), the $s$ integral is trivially
performed by definition of the Dirac $\d$-function, and one obtains
\begin{align}
{\cA}(k^2; p^2,q^2;m^2) &= \int_0^\infty \frac{ds}{k^2 + s}\, \Big(2\pq\, \r_1 + p^2 \r_2 + q^2 \r_5\Big)\nn
=&\int_0^\infty ds\, \r_1(s; p^2,q^2;m^2)
- \frac{e^2\,m^2}{\p^2} \int_0^1 dx \int_0^{1-x} dy \ \frac{\,1}{D}
\label{intrho1}
\end{align}
Comparing (\ref{intrho1}) to (\ref{axdivA}) we have finally that
\be
\int_0^\infty ds\, \r_1(s; p^2,q^2;m^2) = \int_0^\infty ds\, \r_1(s; p^2,q^2;m^2\!=\!0)
= {\cA}(k^2; p^2,q^2;m^2\!=\!0) = \frac{e^2}{2 \p^2}
\label{sumrule}
\ee
independently of $m$ (and also of $p^2, q^2 \ge 0$), which can be verified directly from (\ref{rho1}).

Thus the axial anomaly in the real or dispersive part of the triangle amplitude at $m=0$, (\ref{A0}) is associated
with the {\it UV finite} spectral sum rule (\ref{sumrule}) which holds {\it for all} $p^2,q^2, m^2 > 0$, and also
in the limiting case where $p^2= q^2 =0$ and $m^2 \to 0^+$. This is possible despite the fact that $\r_1(s; p^2,q^2;m^2)$
vanishes pointwise as $p^2, q^2, m^2 \to 0^+$ (in any order) for any non-zero $s$, because
\be
\lim_{p^2, q^2 m^2 \to 0^+} \r_1(s; p^2,q^2;m^2) = \frac{e^2}{2\p^2}\, \d(s)
\label{rho1del}
\ee
becomes a singular Dirac $\d$-function in this limit, as may also be verified by inspection of (\ref{rho1}).

The imaginary, absorptive part of the axial current divergence $\cA$ of (\ref{axanom}) or (\ref{sumrhoi})
vanishes for $m=0$, in accordance with expectations from the classical theory, since $(k^2 + s) \r_1 (s)\big\vert_{s= -k^2}=0$.
However computing the real, dispersive part of $k_\m\G_5^{\m \a\b}$ requires dividing (\ref{sumrhoi}) by $k^2 +s$
(the relativistic analog of the energy denominator of first order perturbation theory) which cancels the first factor,
leaving $\r_1(s)$ itself. By direct calculation the spectral function $\r_1(s)$ does not vanish, even in the $m \to 0$ limit.
The reason for this can be traced to the fact that even in the massless limit, a charged Dirac fermion can make a transition
to a virtual state of the opposite helicity by the emission of a photon~\cite{LeeNau:1964,DolZak:1971,Huang}, that is to say,
massless Dirac fermions do {\it not} become two decoupled Weyl fermions in the presence of electromagnetism. This
results in the matrix elements $\lag n| J^\a J^\b|0 \rag$ of the two electromagnetic current vertices to the fermion
pair intermediate state in the imaginary part of the triangle diagram Fig.~\ref{Fig:J5JJ} remaining non-vanishing
in the massless fermion limit, and contributing positively to $\r_1(s;p^2, q^2;m^2\!=\!0)$ as in (\ref{rho1}).
That its integral over $s$, as in (\ref{sumrule}), gives a finite constant result independent of $p^2, q^2 \ge 0$
is a statement of the finite constant residue of the $1/k^2$ anomaly pole for $m=0$ in the longitudinal part
of the triangle amplitude, protected by the Adler--Bardeen theorem~\cite{AdlerBard:1969}.

It is significant that this dispersive analysis, relying only upon the well-defined, finite imaginary
absorptive parts of the triangle amplitude $\G_5^{\m\a\b}$, matrix elements of low energy states,
and {\it unsubtracted} Kramers--Kronig dispersion relations (\ref{KraKro}), leads to the full amplitude
obeying conditions \ref{inva}-\ref{invd} of Sec.~\ref{Sec:TriAnom}, and the axial anomaly (\ref{divJ5}),
without encountering UV divergent integrals requiring regularization at any step. The appearance of the
$1/k^2$ gapless pole singularity with residue fixed by the finite sum rule (\ref{sumrule}) demonstrate
that the axial anomaly is an infrared (more properly, relativistic lightcone) feature of the low energy effective
theory, independent of UV short distance physics~\cite{DolZak:1971, Horejsi:1986, GiaEM:2009}.

\section{The Axial Anomaly for WSMs in a Constant Uniform Magnetic Field}
\label{App:DimRed}

In a constant uniform magnetic field $\vec B\! = B \bf \hat x$, the Fourier transform of the
electromagnetic potential $\tilde A_{\g =2,3}(q)$ is non-zero in the transverse $\bf \hat y$,
$\!\bf \hat z$ directions, while the constancy of $B=F_{23}$ implies that
\be
q_2 \tilde A_3(q) - q_3 \tilde A_2(q) = \int d^{\lsp 4} z\, e^{-i q\cdot z} (q_2 A_3(z) - q_3 A_2(z) ) =- i B \,(2\p)^4\, \d^{\lsp 4}(q)
\ee
proportional to a $\d$-function at $q^\m=0$. The vectors $k^\m = p^\m + q^\m = p^\m$ of Fig.~\ref{Fig:J5JJ} are then equal to each other. If their spatial components are taken to lie also in the $\bf \hat x$ direction, parallel to $\vec B$, then the $\t_2$ tensor in Table \ref{chitens} is the only one contributing to $\tilde\G_5^{\m\a\g}(p,q)\tilde A_{\g}(q)$, which for $\g=2,3$ is non-zero only for $\m = a, \a=b$ ranging over the $0,1$ components.  We then obtain
\be
\int \frac{d^{\lsp 4}q}{(2\p)^4}\,
\G_5^{\,a b\g}(p,q)A_{\g}(q)\Big\vert_{\vec k_\perp = \vec p_\perp=0}
= i B f_2(k^2;k^2,0) \big(k^2 \e^{ab} - k^bk^c\lsp \e^a{\!}_{c}\big)
\label{G5Bint}
\ee
for the constant, uniform $\vec B$ field, in terms of the anti-symmetric tensor $\e^{ab}\equiv \e^{ab23}$ of the remaining two $a, b= 0,1$ spacetime indices. Next noting from (\ref{denom}) that $D= p^2 x(1-x) = k^2 x(1-x)$ for $m=0$ and $q=0$, the function $f_2$ evaluates simply to
 \be
f_2(k^2;k^2,0)\Big\vert_{m=0} = 2f_1(k^2;k^2,0)\Big\vert_{m=0} = \frac{e^2}{2\p^2}\,\frac{1}{k^2}
= \frac{2 \a}{\p}\,\frac{1}{k^2}\,,
\label{f22D}
\ee
explicitly showing the $1/k^2$ pole in this case. Thus the full anomaly vertex $\G_5^{\,a b\g}(p,q)$
is given entirely by its longitudinal projection in this case of constant, uniform $\vec B$.

\begin{figure}[ht]
\centering
\begin{tikzpicture}[decoration={snake, segment length=2.06mm, amplitude=0.7mm}]
    \tikzstyle{arr}=[decoration={markings,mark=at position 1 with
      {\arrow[scale=1.5]{>}}}, postaction={decorate}];
    \tikzstyle{arrmid}=[decoration={markings,mark=at position 0.5 with
      {\arrow[scale=1.5]{<}}}, postaction={decorate}]
    \draw[arrmid] (2,0) arc (0:180:1);
    \draw[arrmid] (0,0) arc (180:360:1);
    \draw[decorate] (2,0)--(3.5,0);
    \draw[arr] (-1.5,0)--(-1,0) node[above, midway] {$k$};
    \draw[arr] (2.5,0.25)--(3,0.25) node[above, midway] {$k$};
    \filldraw[cross,fill=white,thick] (0,0) circle (4pt) node[left=2pt] {$\tj^{\lsp a}$};
    \filldraw[fill=black] (2,0) circle (1.5pt) node[below right] {$\!\!j^b$};
\end{tikzpicture}
\caption{The one-loop axial anomaly in two dimensions.}
\label{Fig:JJ2D}
\end{figure}
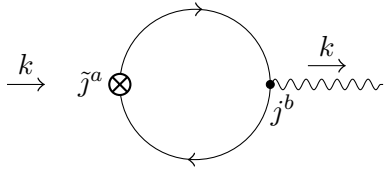

In fact this result for the (three-point) triangle amplitude is closely related to the
(two-point) polarization tensor
\be
\Pi_2^{ab}(k) = i\lag j^a j^b\rag = \frac{1}{\pi k^2} (k^ak^b - k^2 \h^{ab})
\label{Pi2}
\ee
of massless fermions in $1\!+\!1$ dimensional QED$_2$. This 2D polarization tensor (\ref{Pi2}) satisfies
the usual vector WT identity $k_a\lsp\Pi_2^{ab}(k) = 0$ of electric current conservation. On the other hand
the 2D chiral current is
\be
\tj^{\, a} = - \e^a{\!}_{\lsp b}\, j^b\,,
\label{j52D}
\ee
which is dual to the electric current $j^b$.
Thus the chiral current polarization
\be
\widetilde \P_2^{ab}(k) \equiv i \lag \tj^{\lsp a} j^b\rag =-\e^a{\!}_{c}\P_2^{cb}(k)  =\frac{1}{\p k^2}(k^2 \e^{ab} - k^bk^c \e^a{\!}_{c})
\label{Pi2ch}
\ee
satisfies the anomalous WT identity
\be
k_a\widetilde \P_2^{ab}(k) = -\frac{ k_a \e^{ab}}{\p}
\ee
which is the axial anomaly
\be
\pa^a \tj_{a}= \mA_2 = \frac{1}{2\p}\lsp \e^{ab}F_{ab} = \frac{1}{\p}\lsp E
\label{anom2}
\ee
in $1+1$ dimensions, illustrated in Fig.~\ref{Fig:JJ2D}. Here $E=F_{10}$ the electric field strength in
the $\bf \hat x$ direction parallel to $\vec B$.

The dimensional reduction to $1+1$ dimensions results from the fact that with (\ref{f22D}) and (\ref{Pi2ch})
(\ref{G5Bint}) may be written as
\be
\int \frac{d^{\lsp 4}q}{(2\p)^4}\,\G_5^{ab\g}(p,q) A_{\g}(q)\Big\vert_{\vec k_\perp = \vec p_\perp=0}= 2i\a B\,
\lsp\widetilde \P_2^{(5)\,ab}(k)\,,
\label{G5B2D}
\ee
which is non-zero only for $\m=a, \a=b= 0,1$ in a constant uninform magnetic field $\vec B= B \bf \hat x$ in the
indicated limits. Thus the three-point axial triangle anomaly for massless fermions in $3+1$ dimensions becomes proportional
to the two-point polarization tensor $\widetilde \P_2^{ab}$ the 4D axial anomaly becomes proportional
to the 2D axial anomaly, {\it i.e.}
\be
\mA_4\big\vert_B =\frac{eB}{2 \p} \mA_2 = \frac{2\a E B}{\p} \to 2 \a B\,\big(E + \pa_a b^a\big) \,,
\label{anom42}
\ee
after rescaling the electric field $E \to eE$ parallel to $\vec B$, in order to accord with standard (Heaviside--Lorentz) conventions in 4D electromagnetism. In the last replacement of (\ref{anom42}) we have allowed
for the fact that in 2D the chiral potential $b_a = \e_a{\!}^{c}A_c$ is also dual to the electric potential,
{\it i.e.}\ $b_0=A_1, b_1=A_0$, and $\e^{ab}\pa_a A_b \to \e^{ab}A_b + b^a$, so that $\mA_2$ is given by (\ref{anom2}) when both the electric and axial potentials are non-vanishing.

A consequence of this dimensional reduction from 4D to 2D is that the gapless collective boson composed of fermion/anti-fermion pairs propagating along the $\vec B$ direction,  described by the $1/k^2$ anomaly pole in 4D, becomes in this case for ${\bf k}_\perp =0$, precisely the $1/k^2$ propagator pole of the effective boson of the 2D Schwinger model for vanishing 2D electric coupling constant~\cite{Schwinger:1962tp,Blaschke:2014,MotSad:2021}.

The dimensional reduction (\ref{G5B2D}) from 4D to 2D for semimetals in a constant, uniform magnetic field
amounts to the LLL approximation for the $3+1$ dimensional axial polarization tensor
\be
i\big\lag \cT J_5^a(t,x,\by) J^b(t,x',\by')\big\rag_{_{\!\!LLL}}=
2i\a B \int\frac{d\w}{2\pi} \int\frac{dk}{2\pi}\int\frac{d^{\lsp 2}{\bf k}_\perp}{(2\pi)^2}
\,e^{ik\cdot (x-x') + i{\bf k}_\perp \cdot (\by - \by')}e^ {-k_\perp^2/2eB}\lsp\widetilde \P_2^{ab}(\w, k)
\label{LLL}
\ee
in terms of the $1+1$ dimensional one (\ref{Pi2}), along the magnetic field direction \cite{Gusynin:1994xp, Gusynin:1994re, Gusynin:1998zq, Fukushima:2011nu, Miransky:2015ava}. Integrating over the transverse  $\by$ direction, the exponential factor in (\ref{LLL}) is set to unity, so that the Fourier transform in $t,x$ gives
\begin{align}
i \int d^{\lsp 2} x \, e^{-ik\cdot (x-x')} \int d^{\lsp 2}\by\, \big\lag \cT J_5^a(t,x,\by) J^b(t,x',\by')\big\rag_{_\text{LLL}} &= 2i\a B\, \widetilde\P_2^{ab}(k)\nn
&=\int \frac{d^{\lsp 4} q}{(2\p)^4}\,\G_5^{ab\g}(p,q) A_{\g}(q)\Big\vert_{\vec k_\perp = \vec p_\perp=0}
\label{LLL2}
\end{align}
by (\ref{G5B2D}). Thus the axial anomaly of the 4D triangle diagram coincides with the 2D chiral polarization of gapless fermions in a constant, uniform magnetic field in the LLL approximation. The 4D axial current along the magnetic field direction $J_5^a = (e B/2\p) j_5^a, a=0, 1$ in terms of the 2D chiral current $j_5^a$ since $eB/2\p$ is just the electron number density per unit area in the LLL.

The relation (\ref{LLL2}) may also be understood as an expression of the low energy or infrared nature  of the axial anomaly. Since the higher Landau levels are effectively gapped by $eB$, only the LLL's  gapless excitations contribute to the excitations parallel to the $\vec B$ direction at distance  scales much larger than $1/\sqrt{eB}$. If an external electric field is turned on adiabatically, only fermions in the LLL can be excited, and the $d=4$ axial anomaly factorizes into its $d=2$ counterpart  with a transverse density proportional to the uniform magnetic field strength $B$ \cite{Nielsen:1983rb, Basar:2012gm}. The contribution of the axial anomaly to the effective action of massless fermions becomes exact in this limit. Thus the 4D triangle anomaly provides a relevant interaction in the  low energy, long distance EFT of gapless fermionic modes in Weyl and Dirac semimetals placed in a constant, uniform $\vec B$ field.

\section{Integrating Out the Fermions in the Limit of Large Fermion Mass}
\label{App:Intoutfer}
Here we will present a derivation of the effective action (of type \ref{EFTii}) in a WSM from first principles. The action we consider is given by (\ref{Lagf}) and (\ref{NJL}). The four-fermion interaction, which is of the NJL type, is the same as the one considered in the condensed-matter literature, e.g.\ \cite{Wang:2012bgb}. Its treatment in the high-energy physics literature is highly-developed due to the interest in the NJL model as a model for QCD mesons; see e.g.~\cite{Eguchi:1976iz, Ebert:1982pk, Ebert:1997fc}. Our computations will use the covariant derivative expansion techniques described in~\cite{Gaillard:1985uh, Cheyette:1987qz, Henning:2014wua, Zhang:2016pja, Ellis:2020ivx, Angelescu:2020yzf, Quevillon:2021sfz, Filoche:2022dxl}. We will focus only on specific terms of interest and not attempt to derive the full effective field theory.

After the introduction of Hubbard--Stratonovich fields $\phi_1$ and $\phi_2$ given by \eqref{scalars}, the action becomes that of (\ref{Lagf}) with zero bare mass along with (\ref{Lint}). The fermions can now be integrated out to obtain the effective action of $\phi_1,\phi_2,A_\mu$ and $b_\mu$ only:
\eqn{S_{\text{eff}}=-i\lsp\lsp\Tr\ln[\gamma^\mu(i\lsp\partial_\mu + eA_\mu+b_\mu\gamma^5)-g\phi_1-ig\phi_2\gamma^5] -\int d^{\lsp 4}x\,\frac{g^2}{G}(\phi_1^2+\phi_2^2)\,,}[actionHEPtrlog]
where $\Tr$ denotes the full trace over coordinate space and matrices. With standard manipulations assuming continuous momentum fermionic states to express $\Tr$, this can be brought to the form
\eqn{S_{\text{eff}}=\int d^{\lsp 4} x\,
\bigg\{i\int\frac{d^{\lsp 4} k}{(2\pi)^4}\tr\sum_{n=1}^\infty\frac{(-1)^{n}}{n}\bigg(\frac{i\slashed{D} +\slashed{b}\gamma^5-g\phi_1-ig\phi_2\gamma^5}{\slashed{k}-m}\bigg)^n
-\frac{1}{G}\big[(g\phi_1+m)^2+g^2\phi_2^2\big]\bigg\}\,,}[actTrtrick]
where $\slashed{D}=\gamma^\mu D_\mu$, $D_\mu=\partial_\mu-ieA_\mu$, tr
denotes the trace over matrices only, and we shifted $g\phi_1\to g\phi_1+m$. Below we will perform explicit evaluations of terms of interest in $S_{\text{eff}}$ diagramatically. Our calculations will capture terms of the effective action expanded in fields and derivatives. Results for more terms than we consider here are included in \cite{Zhang:2016pja, Ellis:2020ivx, Angelescu:2020yzf, Quevillon:2021sfz}.

First of all, for $g\ne0$ the extremization condition $\delta S_{\text{eff}}/\delta\phi_1|_{\phi_1=\phi_2=0,A_\mu=b_\mu=0}=0$ gives
\eqn{\frac{2m}{G}
=i\int\frac{d^{\lsp 4}k}{(2\pi)^4}\,\tr\frac{1}{\slashed{k}-m}=4\lsp m I_1\,,
}[gapEq]
where\foot{To evaluate the integral we Wick rotate by $k^0=ik_E^0$. With our metric conventions, $k^2=k_E^2$.}
\eqn{I_1=-i\int\frac{d^{\lsp 4}k}{(2\pi)^4} \frac{1}{k^2+m^2}\stackrel{\Lambda\gg m}{=} \frac{1}{16\pi^2}\Big(\Lambda^2 -m^2\ln\frac{\Lambda^2}{m^2}\Big)\,.}[intIone]
We use a hard cutoff $\Lambda$ to regularize the loop integral $I_1$. We will be working with $\Lambda\gg m$.\foot{As usual, the cutoff drops out of all physical quantities because $G$ depends on $\Lambda$. For example, if $m$ is measured, then the gap equation \gapEq simply gives us the dependence of the bare coupling $G$ on $\Lambda$, with $m$ fixed at the measured value. Observables depend on the measured $m$ and not the cutoff-dependent $G$.} With the definition $G_c=8\pi^2/\Lambda^2$, \gapEq becomes
\eqn{2m\Big(\llsp\frac{1}{G}-\frac{1}{G_c}\llsp\Big)=-\frac{m^3}{4\pi^2}\ln\frac{\Lambda^2}{m^2}\,.}[gapEqII]
If $G<G_c$, then the only solution of \gapEqII is $m=0$, but if $G>G_c$ we have a solution with $m\neq0$ which signals spontaneous breaking of chiral symmetry. To examine stability of the $m=0$ vacuum we evaluate
\eqn{-\!\left.\frac{\delta^2 S_{\text{eff}}}{\delta\phi_1^2}\right|_{m=\phi_1=\phi_2=0,A_\mu=b_\mu=0}
=\frac{2g^2}{G}-ig^2\int\frac{d^{\lsp 4}
k}{(2\pi)^4}\,\tr\frac{1}{\slashed{k}^2}=2g^2\Big(\frac{1}{G}-\frac{1}{G_c}\Big)\,.}[SecDeriv]
Thus, for $G<G_c$ (resp.\ $G>G_c$) the $m=0$ vacuum is stable (resp.\ unstable). In the remainder of this section we work in the broken phase ($m\neq0$) and focus on terms in the effective action involving $\phi_2$, $A_\mu$ and $b_\mu$.

Quadratic terms in $\phi_2$ with no derivatives are generated by
\eqn{
  \mathscr{L}_{\phi_2^2}=\begin{tikzpicture}[baseline=(vert_cent.base)]
    \node (vert_cent) {$\phantom{\cdot}$};
    \draw (0.5,0) circle (0.5);
    \filldraw (0,0) circle (1pt) node[left] {$\phi_2$};
    \filldraw (1,0) circle (1pt) node[right] {$\phi_2$};
  \end{tikzpicture}
  = 2g^2I_1\lsp \phi_2^2\,,
}[]
and a kinetic term for $\phi_2$ is generated from the sum of diagrams
\eqn{\mathscr{L}_{D^2\phi_2^2}=
  \begin{tikzpicture}[baseline=(vert_cent.base)]
    \node (vert_cent) {$\phantom{\cdot}$};
    \draw (0,0) circle (0.5);
    \filldraw (45:0.5) circle (1pt) node[above right=-3pt] {$\slashed{D}$};
    \filldraw (135:0.5) circle (1pt) node[above left=-2pt] {$\phi_2$};
    \filldraw (225:0.5) circle (1pt) node[below left=-3pt] {$\slashed{D}$};
    \filldraw (315:0.5) circle (1pt) node[below right=-3pt] {$\phi_2$};
\end{tikzpicture}+
\begin{tikzpicture}[baseline=(vert_cent.base)]
    \node (vert_cent) {$\phantom{\cdot}$};
    \draw (0,0) circle (0.5);
    \filldraw (45:0.5) circle (1pt) node[above right=-2pt] {$\phi_2$};
    \filldraw (135:0.5) circle (1pt) node[above left=-4pt] {$\slashed{D}$};
    \filldraw (225:0.5) circle (1pt) node[below left=-3pt] {$\phi_2$};
    \filldraw (315:0.5) circle (1pt) node[below right=-4pt] {$\slashed{D}$};
\end{tikzpicture}+
\begin{tikzpicture}[baseline=(vert_cent.base)]
    \node (vert_cent) {$\phantom{\cdot}$};
    \draw (0,0) circle (0.5);
    \filldraw (45:0.5) circle (1pt) node[above right=-2pt] {$\phi_2$};
    \filldraw (135:0.5) circle (1pt) node[above left=-2pt] {$\phi_2$};
    \filldraw (225:0.5) circle (1pt) node[below left=-3pt] {$\slashed{D}$};
    \filldraw (315:0.5) circle (1pt) node[below right=-4pt] {$\slashed{D}$};
\end{tikzpicture}+
\begin{tikzpicture}[baseline=(vert_cent.base)]
    \node (vert_cent) {$\phantom{\cdot}$};
    \draw (0,0) circle (0.5);
    \filldraw (45:0.5) circle (1pt) node[above right=-3pt] {$\slashed{D}$};
    \filldraw (135:0.5) circle (1pt) node[above left=-2pt] {$\phi_2$};
    \filldraw (225:0.5) circle (1pt) node[below left=-3pt] {$\phi_2$};
    \filldraw (315:0.5) circle (1pt) node[below right=-4pt] {$\slashed{D}$};
  \end{tikzpicture}+
  \begin{tikzpicture}[baseline=(vert_cent.base)]
    \node (vert_cent) {$\phantom{\cdot}$};
    \draw (0,0) circle (0.5);
    \filldraw (45:0.5) circle (1pt) node[above right=-3pt] {$\slashed{D}$};
    \filldraw (135:0.5) circle (1pt) node[above left=-4pt] {$\slashed{D}$};
    \filldraw (225:0.5) circle (1pt) node[below left=-3pt] {$\phi_2$};
    \filldraw (315:0.5) circle (1pt) node[below right=-3pt] {$\phi_2$};
\end{tikzpicture}+
\begin{tikzpicture}[baseline=(vert_cent.base)]
    \node (vert_cent) {$\phantom{\cdot}$};
    \draw (0,0) circle (0.5);
    \filldraw (45:0.5) circle (1pt) node[above right=-2pt] {$\phi_2$};
    \filldraw (135:0.5) circle (1pt) node[above left=-4pt] {$\slashed{D}$};
    \filldraw (225:0.5) circle (1pt) node[below left=-3pt] {$\slashed{D}$};
    \filldraw (315:0.5) circle (1pt) node[below right=-2pt] {$\phi_2$};
\end{tikzpicture}\,.
}[LagDsqpisq]
Due to cyclicity of the trace\foot{Here we invoke cyclicity of Tr as opposed to tr; see~\cite[Sec.\ 3.4]{Angelescu:2020yzf}.} the first two diagrams in the right-hand side of \eqref{LagDsqpisq} are equal, as are the second four.  We have
\eqn{
  \begin{tikzpicture}[baseline=(vert_cent.base)]
    \node (vert_cent) {$\phantom{\cdot}$};
    \draw (0,0) circle (0.5);
    \filldraw (45:0.5) circle (1pt) node[above right=-3pt] {$\slashed{D}$};
    \filldraw (135:0.5) circle (1pt) node[above left=-2pt] {$\phi_2$};
    \filldraw (225:0.5) circle (1pt) node[below left=-3pt] {$\slashed{D}$};
    \filldraw (315:0.5) circle (1pt) node[below right=-3pt] {$\phi_2$};
\end{tikzpicture}=-g^2I_2\lsp D^\mu\phi_2 D_\mu\phi_2\qquad\text{and}\qquad
\begin{tikzpicture}[baseline=(vert_cent.base)]
    \node (vert_cent) {$\phantom{\cdot}$};
    \draw (0,0) circle (0.5);
    \filldraw (45:0.5) circle (1pt) node[above right=-2pt] {$\phi_2$};
    \filldraw (135:0.5) circle (1pt) node[above left=-2pt] {$\phi_2$};
    \filldraw (225:0.5) circle (1pt) node[below left=-3pt] {$\slashed{D}$};
    \filldraw (315:0.5) circle (1pt) node[below right=-4pt] {$\slashed{D}$};
\end{tikzpicture}
=g^2I_2\lsp\phi_2^2 D^2-2g^2J_2^{\mu\nu}\lsp\phi_2^2 D_\mu D_\nu\,,}[]
where
\eqn{I_2=-i\int\frac{d^{\lsp 4}k}{(2\pi)^4}\frac{1}{(k^2+m^2)^2}\stackrel{\Lambda\gg m}{=}-\frac{1}{16\pi^2}\Big(1-\ln\frac{\Lambda^2}{m^2}\Big)\,,}[intItwo]
\eqn{J_2^{\mu\nu}=-i\int\frac{d^{\lsp 4}k}{(2\pi)^4} \frac{k^\mu k^\nu}{(k^2+m^2)^3}\,.}[intJ2munu]
With a gauge invariant regulator or the introduction of suitable counterterms we may obtain $J_2^{\mu\nu}=\frac14\eta^{\mu\nu}I_2$. Then,
\eqna{\mathscr{L}_{D^2\phi_2^2}=-g^2I_2(2\lsp D^\mu\phi_2 D_\mu\phi_2-2\lsp \phi_2^2 D^2)=-g^2I_2\lsp[D^\mu,\phi_2][D_\mu,\phi_2]\,.}[LagDsqpisqII]
In our case, $[D_\mu,\phi_2]=\partial_\mu\phi_2$. As a result, in the effective action we will have the terms
\eqn{S_{\phi_2^2}+S_{\partial^2\phi_2^2}= \int d^{\lsp 4}x\,
(\mathscr{L}_{\phi_2^2}+\mathscr{L}_{\partial^2\phi_2^2})
=\int d^{\lsp 4}x\,\big[\!-\!g^2\big(\tfrac{1}{G}-2\lsp I_1\big)\lsp\phi_2^2
- g^2I_2\lsp
\partial^\mu\phi_2\lsp\partial_\mu\phi_2\big]\,.}[quadraticPi]
The $\phi_2^2$ term is eliminated by the gap equation \gapEq and we are left with
\eqn{S_{\partial^2\phi_2^2}= -\int d^{\lsp 4}x\,g^2I_2\lsp
\partial^\mu\phi_2 \lsp\partial_\mu\phi_2\,,}[kinpi]
which is the kinetic term for the massless Nambu--Goldstone boson of spontaneous chiral symmetry breaking.

There are no terms linear in $\phi_2$ and $b$ with no derivatives because the corresponding trace is zero, but with one derivative in addition we have the sum of diagrams
\eqn{\mathscr{L}_{D\phi_2 b}=\begin{tikzpicture}[baseline=(vert_cent.base)]
    \node (vert_cent) {$\phantom{\cdot}$};
    \draw (0,0) circle (0.5);
    \filldraw (300:0.5) circle (1pt) node[below right=-2pt] {$\slashed{b}$};
    \filldraw (180:0.5) circle (1pt) node[left] {$\phi_2$};
    \filldraw (60:0.5) circle (1pt) node[above right=-2pt]
      {$\slashed{D}$};
\end{tikzpicture}+
\begin{tikzpicture}[baseline=(vert_cent.base)]
    \node (vert_cent) {$\phantom{\cdot}$};
    \draw (0,0) circle (0.5);
    \filldraw (300:0.5) circle (1pt) node[below right=-2pt] {$\phi_2$};
    \filldraw (180:0.5) circle (1pt) node[left] {$\slashed{D}$};
    \filldraw (60:0.5) circle (1pt) node[above right=-2pt]
      {$\slashed{b}$};
\end{tikzpicture}+
\begin{tikzpicture}[baseline=(vert_cent.base)]
    \node (vert_cent) {$\phantom{\cdot}$};
    \draw (0,0) circle (0.5);
    \filldraw (300:0.5) circle (1pt) node[below right=-3pt]
      {$\slashed{D}$};
    \filldraw (180:0.5) circle (1pt) node[left] {$\slashed{b}$};
    \filldraw (60:0.5) circle (1pt) node[above right=-2pt] {$\phi_2$};
\end{tikzpicture}+
\begin{tikzpicture}[baseline=(vert_cent.base)]
    \node (vert_cent) {$\phantom{\cdot}$};
    \draw (0,0) circle (0.5);
    \filldraw (300:0.5) circle (1pt) node[below right=-2pt] {$\phi_2$};
    \filldraw (180:0.5) circle (1pt) node[left] {$\slashed{b}$};
    \filldraw (60:0.5) circle (1pt) node[above right=-2pt]
      {$\slashed{D}$};
\end{tikzpicture}+
\begin{tikzpicture}[baseline=(vert_cent.base)]
    \node (vert_cent) {$\phantom{\cdot}$};
    \draw (0,0) circle (0.5);
    \filldraw (300:0.5) circle (1pt) node[below right=-2pt] {$\slashed{b}$};
    \filldraw (180:0.5) circle (1pt) node[left] {$\slashed{D}$};
    \filldraw (60:0.5) circle (1pt) node[above right=-2pt] {$\phi_2$};
  \end{tikzpicture}+
\begin{tikzpicture}[baseline=(vert_cent.base)]
    \node (vert_cent) {$\phantom{\cdot}$};
    \draw (0,0) circle (0.5);
    \filldraw (300:0.5) circle (1pt) node[below right=-3pt]
      {$\slashed{D}$};
    \filldraw (180:0.5) circle (1pt) node[left] {$\phi_2$};
    \filldraw (60:0.5) circle (1pt) node[above right=-2pt]
      {$\slashed{b}$};
\end{tikzpicture}\,.}[]
The first three diagrams are equal, as are the last three. A straightforward calculation gives
\eqn{S_{\partial\phi_2 b}=-\int d^{\lsp 4}x\,4g\lsp m I_2
\lsp\partial^\mu\phi_2\lsp b_\mu\,.}[etaAfivemixed]

Further terms of relevance to our discussion arise at fifth order. More specifically, we are interested in terms with one power of $\phi_2$ and four derivatives:
\eqn{\mathscr{L}_{D^4\phi_2}=\begin{tikzpicture}[baseline=(vert_cent.base)]
    \node (vert_cent) {$\phantom{\cdot}$};
    \draw (0,0) circle (0.5);
    \filldraw (324:0.5) circle (1pt) node[below right=-3pt] {$\slashed{D}$};
    \filldraw (252:0.5) circle (1pt) node[below left=-2pt]
      {$\slashed{D}$};
    \filldraw (180:0.5) circle (1pt) node[left] {$\slashed{D}$};
    \filldraw (108:0.5) circle (1pt) node[above left=-3pt]
      {$\slashed{D}$};
    \filldraw (36:0.5) circle (1pt) node[above right=-2pt] {$\phi_2$};
  \end{tikzpicture}\!+\cdots=-\frac{4g}{m}I_3\lsp
  \epsilon^{\kappa\lambda\mu\nu}\phi_2\lsp
  D_\kappa D_\lambda D_\mu D_\nu=
  -\frac{g}{m}I_3\lsp\epsilon^{\kappa\lambda\mu\nu}\phi_2\lsp
  [D_\kappa,D_\lambda][D_\mu,D_\nu]\,,
}[anompi]
where
\eqn{I_3=-im^2\int\frac{d^{\lsp 4}k}{(2\pi)^4}\frac{1}{(k^2+m^2)^3} \stackrel{\Lambda\gg m}{=} \frac{1}{32\pi^2}\,.}[intIthree]
Since $[D_\mu,D_\nu]=ieF_{\mu\nu}$, we find
\eqn{S_{\phi_2\partial^2 A^2}=\int d^{\lsp 4}x\,\frac{2g\lsp e^2}{m}I_3
\lsp\phi_2\lsp F_{\mu\nu}\widetilde{F}^{\mu\nu}=
\frac{g\lsp e^2}{16\pi^2}\int d^{\lsp 4}x\,
\frac{\phi_2}{m}F_{\mu\nu}\widetilde{F}^{\mu\nu}\,,}[etaonaxion]

Collecting the contributions \kinpi, \etaAfivemixed and \etaonaxion, we have
\eqn{S_{\text{eff}}\supset\int d^{\lsp 4}x\,\Big(
\!-\!g^2I_2\lsp\partial^\mu\phi_2 \lsp\partial_\mu\phi_2
-4g\lsp m I_2\lsp\partial^\mu\phi_2\lsp b_\mu
+\frac{g\lsp e^2}{16\pi^2}\frac{\phi_2}{m}F_{\mu\nu}\widetilde{F}^{\mu\nu}
\Big)\,.}[actsofarI]
Further terms involving derivatives of $b_\mu$ should be included, e.g.\ an $F^5_{\mu\nu}\widetilde{F}^{5\lsp\mu\nu}$ term with $F^5_{\mu\nu}=\partial_\mu b_\nu-\partial_\nu b_\mu$ (see \cite{Quevillon:2021sfz}), but we omit them here.

Terms quadratic in $A$ and terms quadratic in $b$ are given respectively
by
\eqn{
  \mathscr{L}_{D^2}=\begin{tikzpicture}[baseline=(vert_cent.base)]
    \node (vert_cent) {$\phantom{\cdot}$};
    \draw (0.5,0) circle (0.5);
    \filldraw (0,0) circle (1pt) node[left] {$\slashed{D}$};
    \filldraw (1,0) circle (1pt) node[right] {$\slashed{D}$};
  \end{tikzpicture}
  =-2\lsp I_1\lsp D^2 +4\lsp J_1^{\mu\nu}D_\mu D_\nu\,,
}[DDcont]
where
\eqn{J_1^{\mu\nu}=-i\int \frac{d^{\lsp 4}k}{(2\pi)^4}\frac{k^\mu k^\nu}{(k^2+m^2)^2}\,,}[intJ1munu]
and
\eqn{
  \mathscr{L}_{b^2}
  =\begin{tikzpicture}[baseline=(vert_cent.base)]
    \node (vert_cent) {$\phantom{\cdot}$};
    \draw (0.5,0) circle (0.5);
    \filldraw (0,0) circle (1pt) node[left] {$\slashed{b}$};
    \filldraw (1,0) circle (1pt) node[right] {$\slashed{b}$};
  \end{tikzpicture}
  =-(4\lsp m^2I_2\lsp \eta^{\mu\nu}+2\lsp I_1\lsp \eta^{\mu\nu} -4\lsp J_1^{\mu\nu})\lsp b_\mu b_\nu\,.
}[AfiveAfiveCont]
To preserve gauge invariance we need to demand $\mathscr{L}_{D^2}=0$, or $J_1^{\mu\nu}=\frac12\eta^{\mu\nu}I_1$. With this choice we still get a mass for the axial vector.
Adding this to \actsofarI we get
\eqn{S_{\text{eff}}\supset\int d^{\lsp 4}x\,\Big(\!-\!4\lsp m^2 I_2\lsp B^\mu B_\mu
+\frac{g\lsp e^2}{16\pi^2}\frac{\phi_2}{m}F_{\mu\nu}\widetilde{F}^{\mu\nu}\Big)
\,,\qquad B_\mu=b_\mu+\frac{g}{2\lsp m}\partial_\mu\phi_2\,.}[effTh]

An infinitesimal chiral rotation of the original fermion with $x$-dependent parameter $\beta=\beta(x)$, namely
\eqn{\psi\to(1+i\beta\gamma^5)\psi\,,\qquad \psib\to\psib(1+i\beta\gamma^5)\,,}[chirrot]
which results in the appearance of an anomaly term $\int d^{\lsp 4}x\,\frac{e^2}{8\pi^2}\beta\lsp F_{\mu\nu}\widetilde{F}^{\mu\nu}$ in the effective action, leaves the theory \effTh invariant if $g\phi_2\to g\phi_2-2\lsp m\lsp\beta$ and $b_\mu\to b_\mu+\partial_\mu\beta$. This guarantees that \anompi is the only source of the anomaly in this picture.

A quadratic and a quartic contribution in $\phi_1$ are generated via
\eqn{
  \begin{tikzpicture}[baseline=(vert_cent.base)]
    \node (vert_cent) {$\phantom{\cdot}$};
    \draw (0.5,0) circle (0.5);
    \filldraw (0,0) circle (1pt) node[left] {$\phi_1$};
    \filldraw (1,0) circle (1pt) node[right] {$\phi_1$};
  \end{tikzpicture}
  = 2g^2(I_1-2\lsp m^2 I_2)\lsp \phi_1^2\,,\qquad
  \begin{tikzpicture}[baseline=(vert_cent.base)]
    \node (vert_cent) {$\phantom{\cdot}$};
    \draw (0,0) circle (0.5);
    \filldraw (45:0.5) circle (1pt) node[above right=-3pt] {$\phi_1$};
    \filldraw (135:0.5) circle (1pt) node[above left=-2pt] {$\phi_1$};
    \filldraw (225:0.5) circle (1pt) node[below left=-3pt] {$\phi_1$};
    \filldraw (315:0.5) circle (1pt) node[below right=-3pt] {$\phi_1$};
\end{tikzpicture}=g^4(-I_2+8\lsp I_3-8\lsp I_4)\phi_1^4\,,
}[quadquarphi1]
where
\eqn{I_4=-im^4\int\frac{d^{\lsp 4}k}{(2\pi)^4}\frac{1}{(k^2+m^2)^4} \stackrel{\Lambda\gg m}{=} \frac{1}{96\pi^2}\,.}[intIfour]
In the large $\Lambda/m$ limit we find
\eqn{
  \begin{tikzpicture}[baseline=(vert_cent.base)]
    \node (vert_cent) {$\phantom{\cdot}$};
    \draw (0.5,0) circle (0.5);
    \filldraw (0,0) circle (1pt) node[left] {$\phi_1$};
    \filldraw (1,0) circle (1pt) node[right] {$\phi_1$};
  \end{tikzpicture}
  = g^2\frac{\Lambda^2}{8\pi^2}\lsp\phi_1^2=\frac{g^2}{G_c}\lsp\phi_1^2\,,\qquad
  \begin{tikzpicture}[baseline=(vert_cent.base)]
    \node (vert_cent) {$\phantom{\cdot}$};
    \draw (0,0) circle (0.5);
    \filldraw (45:0.5) circle (1pt) node[above right=-3pt] {$\phi_1$};
    \filldraw (135:0.5) circle (1pt) node[above left=-2pt] {$\phi_1$};
    \filldraw (225:0.5) circle (1pt) node[below left=-3pt] {$\phi_1$};
    \filldraw (315:0.5) circle (1pt) node[below right=-3pt] {$\phi_1$};
\end{tikzpicture}=-\frac{g^4}{16\pi^2}\ln\frac{\Lambda^2}{m^2}\,\phi_1^4\,.
}[quadquarphi2]
Thus,
\eqn{\mathscr{L}_{\phi_1^2}+\mathscr{L}_{\phi_1^4}= -\frac12g^2\Big(\frac{2}{G}-\frac{2}{G_c}\Big)\phi_1^2-\frac14\frac{g^4}{4\pi^2}\ln\frac{\Lambda^2}{m^2}\,\phi_1^4\,,}[kaplamcalcEFT]
which is consistent with \SecDeriv. Requiring that
$m^2=-\kappa/\lambda, \kappa<0$, now gives us the gap equation \gapEqII
once again. Comparing with \eqref{Vpot}, we find \eqref{kapNJL} and \eqref{lamNJL}.

Note that the quartic in $\phi_2$ contribution is
\eqn{\mathscr{L}_{\phi_2^4}= \begin{tikzpicture}[baseline=(vert_cent.base)]
    \node (vert_cent) {$\phantom{\cdot}$};
    \draw (0,0) circle (0.5);
    \filldraw (45:0.5) circle (1pt) node[above right=-3pt] {$\phi_2$};
    \filldraw (135:0.5) circle (1pt) node[above left=-2pt] {$\phi_2$};
    \filldraw (225:0.5) circle (1pt) node[below left=-3pt] {$\phi_2$};
    \filldraw (315:0.5) circle (1pt) node[below right=-3pt] {$\phi_2$};
\end{tikzpicture}=-g^4I_2\,\phi_2^4\,,}[]
which is exactly as required for consistency with \eqref{Vpot}, since $\sigma^2=\phi_1^2+\phi_2^2$.

With the polar representation (\ref{Phipolar}), $\phi_1^2+\phi_2^2=\sigma^2$ and the field $\zeta$ is dimensionless.  Using (\ref{Phipolar}) in our action, given by (\ref{Lagf}) with zero bare mass along with (\ref{Lint}), we get
\eqn{S=\int d^{\lsp 4}x\,\Big\{\bar{\psi}\big[\gamma^\mu(
i\lsp\lrpr_\mu+eA_\mu+b_\mu\gamma^5) -g\lsp\sigma e^{2i\zeta\gamma^5}\big]\psi-\frac{g^2}{G}\lsp \sigma^2
\Big\}\,.
}[OrigActNonlin]
To compute the effective action associated with integrating out $\psi$ we define a chirally rotated fermion $\psi'= e^{i\zeta\gamma^5}\psi$. Then,
\eqn{S=\int d^{\lsp 4}x\,\Big\{\bar{\psi}'[\gamma^\mu(
i\lsp\lrpr_\mu+eA_\mu+B\smash{'}{\!\!}_\mu\gamma^5) -g\lsp\sigma]\psi'-\frac{g^2}{G}\lsp \sigma^2+\frac{e^2}{8\pi^2}\lsp\zeta\lsp F_{\mu\nu}\widetilde{F}^{\mu\nu}\Big\}\,,\qquad B\smash{'}{\!\!}_\mu=b_\mu+\partial_\mu\zeta\,,}[anomNonLin]
and the fermion $\psi'$ can again be integrated out as above. It is important to note that the chiral rotation of the fermion results in a contribution to the effective action associated with the transformation of the path-integral measure, namely the last term in \anomNonLin. This contribution has been discussed in~\cite{Filoche:2022dxl} using covariant derivative expansion methods. The linear dependence on $\z$ in \anomNonLin follows from the Adler--Bardeen theorem~\cite{AdlerBard:1969}, together with Bardeen's general results for the axial anomaly~\cite{Bardeen:1969}.

\end{appendices}

\bibliography{alpsbib}

\end{document}